\documentclass{aa}
\usepackage{natbib}         
\usepackage{graphicx,color}
\usepackage{txfonts}

\newcommand{\rmd}{ {\mathrm d} }

\newcommand{\dVxdt}{ (\rmd V_x / \rmd t)_{\rm fit}}

\begin{document}

\authorrunning{Gulisano \textit{et al.}}

\title{Global and local expansion of magnetic clouds in the inner heliosphere}

\author{A.M. Gulisano \inst{1,2}, P. D\'emoulin \inst{3},
S. Dasso \inst{1,2}, M. E. Ruiz \inst{1,2}, and E. Marsch \inst{4}}
\offprints{A.M. Gulisano}
\institute{
$^{1}$ Instituto de Astronom\'\i a y F\'\i sica del Espacio, CONICET-UBA,
CC. 67, Suc. 28, 1428 Buenos Aires, Argentina \email{agulisano@iafe.uba.ar, sdasso@iafe.uba.ar, meruiz@iafe.uba.ar}\\
$^{2}$ Departamento de F\'\i sica, Facultad de Ciencias Exactas y
Naturales, Universidad de Buenos Aires, 1428 Buenos Aires, Argentina\\
$^{3}$ Observatoire de Paris, LESIA, UMR 8109 (CNRS),
       F-92195 Meudon Principal Cedex, France \email{Pascal.Demoulin@obspm.fr}\\
$^{4}$ Max-Planck-Institut f\"ur Sonnensystemforschung, 37191 Katlenburg-Lindau, Germany;  \email{marsch@mps.mpg.de}\\
}
   \date{Received ***; accepted ***}

   \abstract
   {Observations of magnetic clouds (MCs) are consistent with the presence
    of flux ropes detected in the solar wind (SW) a few days after
their expulsion from the Sun as coronal mass ejections (CMEs).}
   {Both the \textit{in situ} observations of plasma velocity profiles and
the increase of their size with solar distance show that MCs are
typically expanding structures. The aim of this work is to derive
the expansion properties of MCs in the inner heliosphere from 0.3 to
1 AU.}
   {We analyze MCs observed by the two Helios spacecraft using \textit{in
   situ} magnetic field and velocity measurements.
We split the sample in two subsets: those MCs with a velocity
profile that is significantly perturbed from the expected linear
profile and those that are not.

From the slope of the \textit{in situ} measured bulk velocity along
the Sun-Earth direction, we compute an expansion speed with respect to the cloud center
for each of the analyzed MCs.}
   {We analyze how the expansion speed 
depends on the MC size, the translation velocity, and the
heliocentric distance, finding that all MCs in the subset of
non-perturbed MCs expand with almost the same non-dimensional
expansion rate ($\zeta$). We find departures from this general rule
for $\zeta$ only for perturbed MCs, and we interpret the departures
as the consequence of a local and strong SW perturbation by SW fast
streams, affecting the MC even inside its interior, in addition to
the direct interaction region between the SW and the MC. We also
compute the dependence of the mean total SW pressure on the solar
distance and we confirm that the decrease of the total SW pressure
with distance is the main origin of the observed MC expansion rate.
We found that $\zeta$ was $0.91\pm 0.23$ for non-perturbed MCs while
$\zeta$ was $0.48\pm 0.79$ for perturbed MCs, the larger spread in
the last ones being due to the influence of the solar wind local
environment conditions on the expansion. }
{}
    \keywords{Sun: magnetic fields, Magnetohydrodynamics (MHD), Sun:
coronal mass ejections (CMEs), Sun: solar wind, Interplanetary
medium }

\maketitle

\section{Introduction} 
\label{Introduction}

Magnetic clouds (MCs) are magnetized plasma structures forming a
particular subset of interplanetary coronal mass ejections
\citep[ICMEs, e.g.,][]{Burlaga95}. MCs are transient structures in
the solar wind (SW) defined by an enhanced magnetic field with
respect to that found in the surrounding SW with a coherent rotation
of the field of the order of about a day when these structures are
observed at 1 AU \citep{Burlaga81}.
 A lower proton temperature than
the expected one in the SW with the same velocity is another
signature of MCs that complement their identification
\citep[e.g.,][]{Richardson95,Marsch09}.

\begin{figure*}[t!]
\centerline{
\includegraphics[width=0.4\textwidth, clip=]{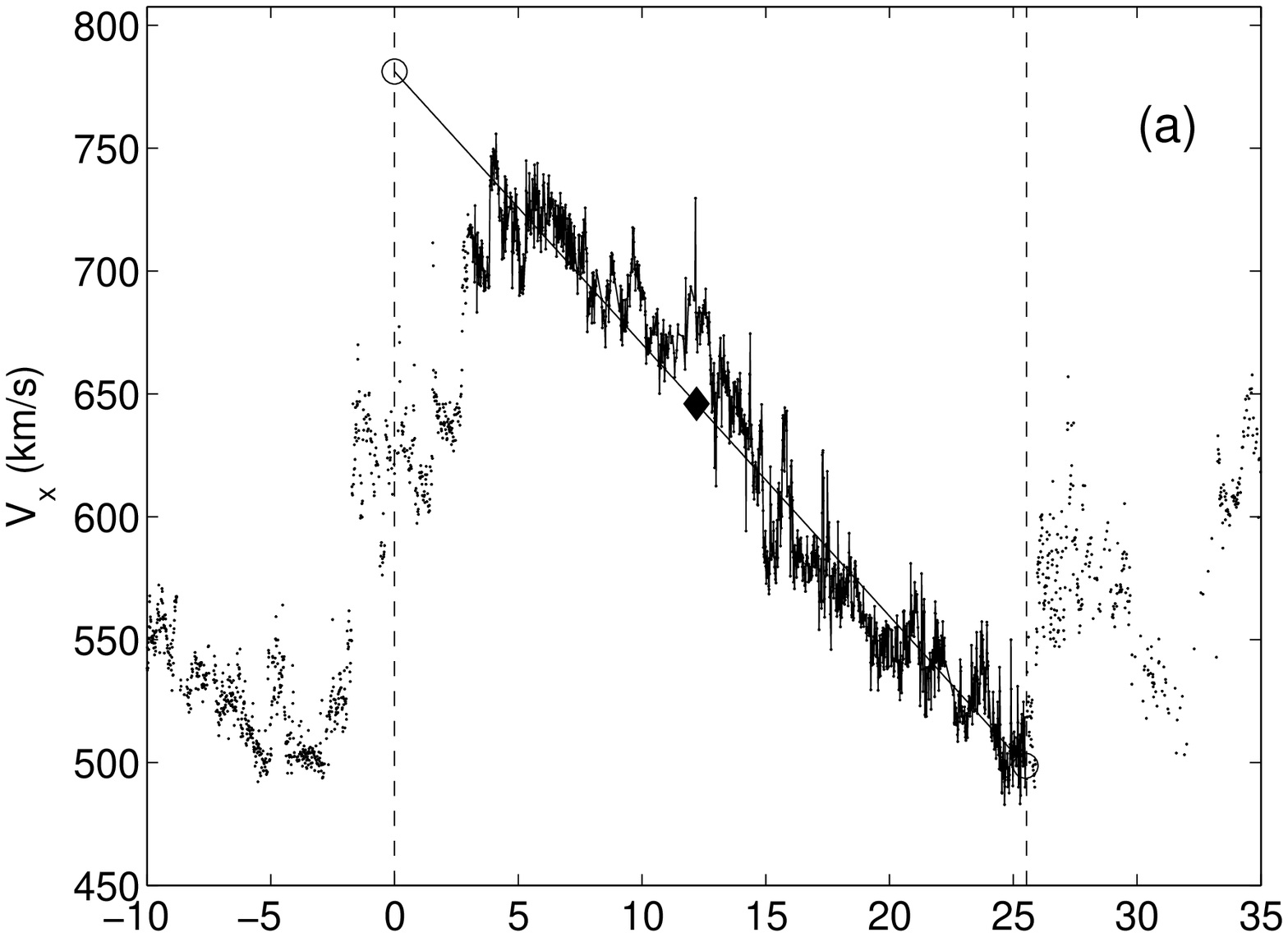}
\includegraphics[width=0.4\textwidth, clip=]{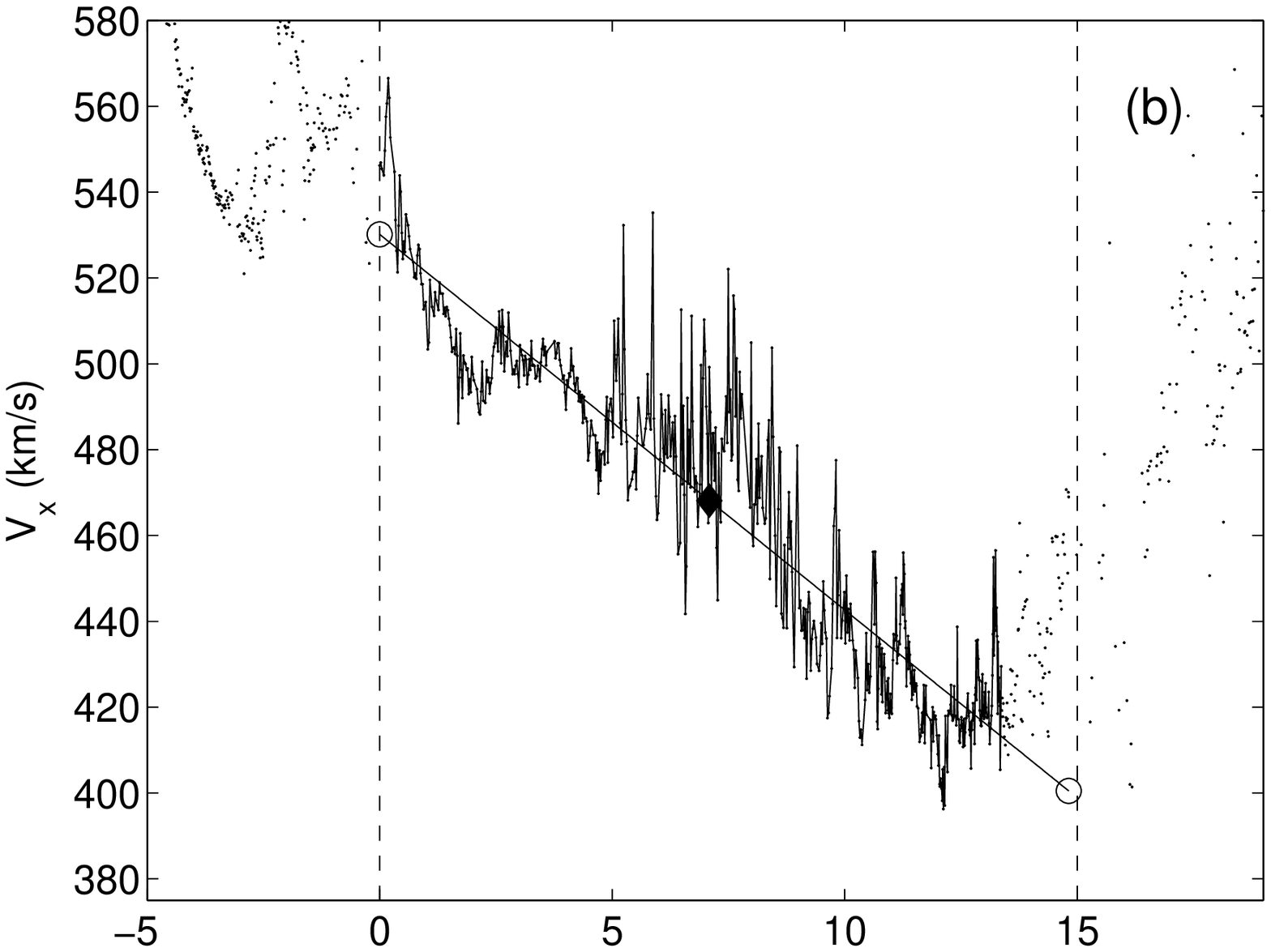}}
\centerline{\includegraphics[width=0.4\textwidth,clip=]{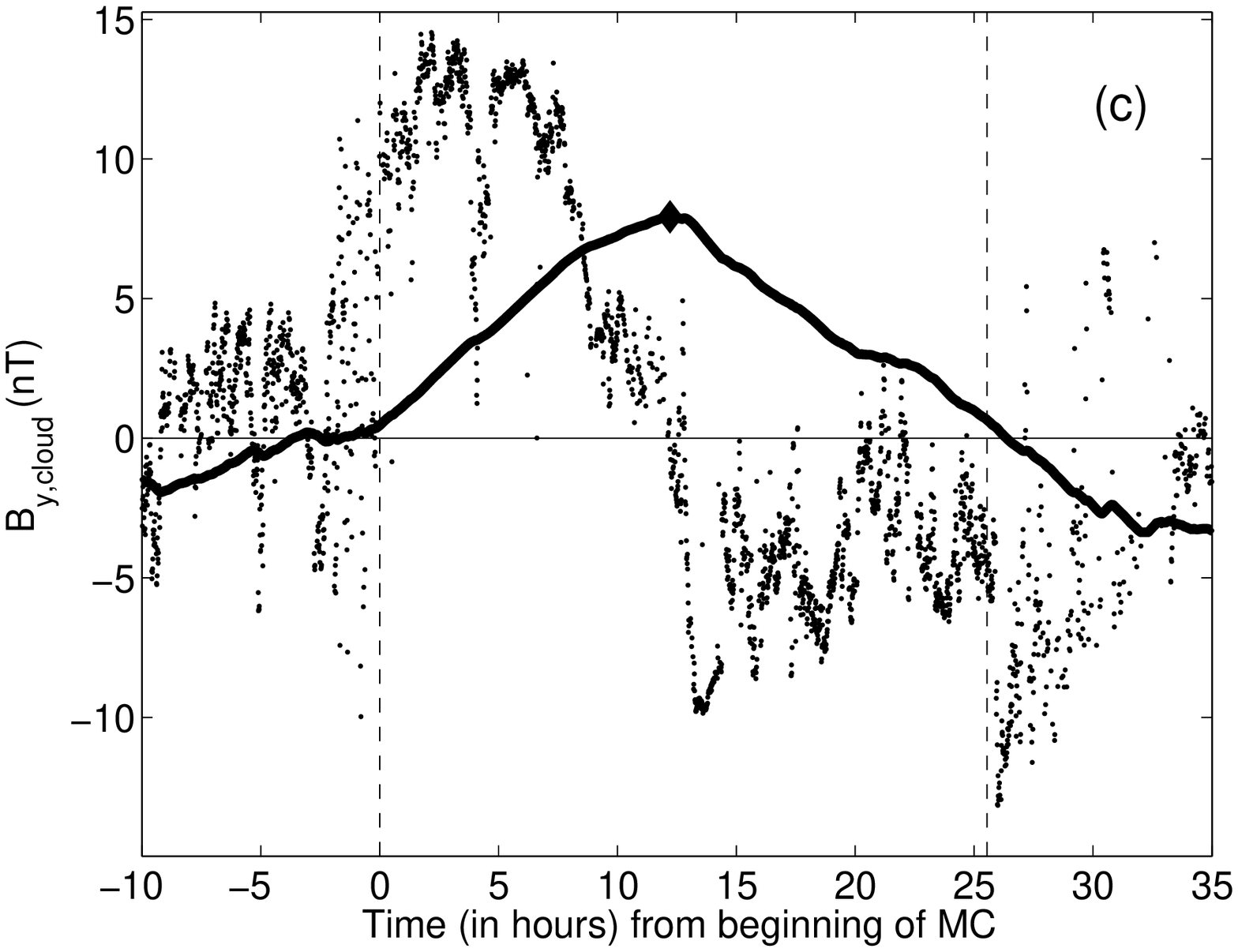}
\includegraphics[width=0.4\textwidth, clip=]{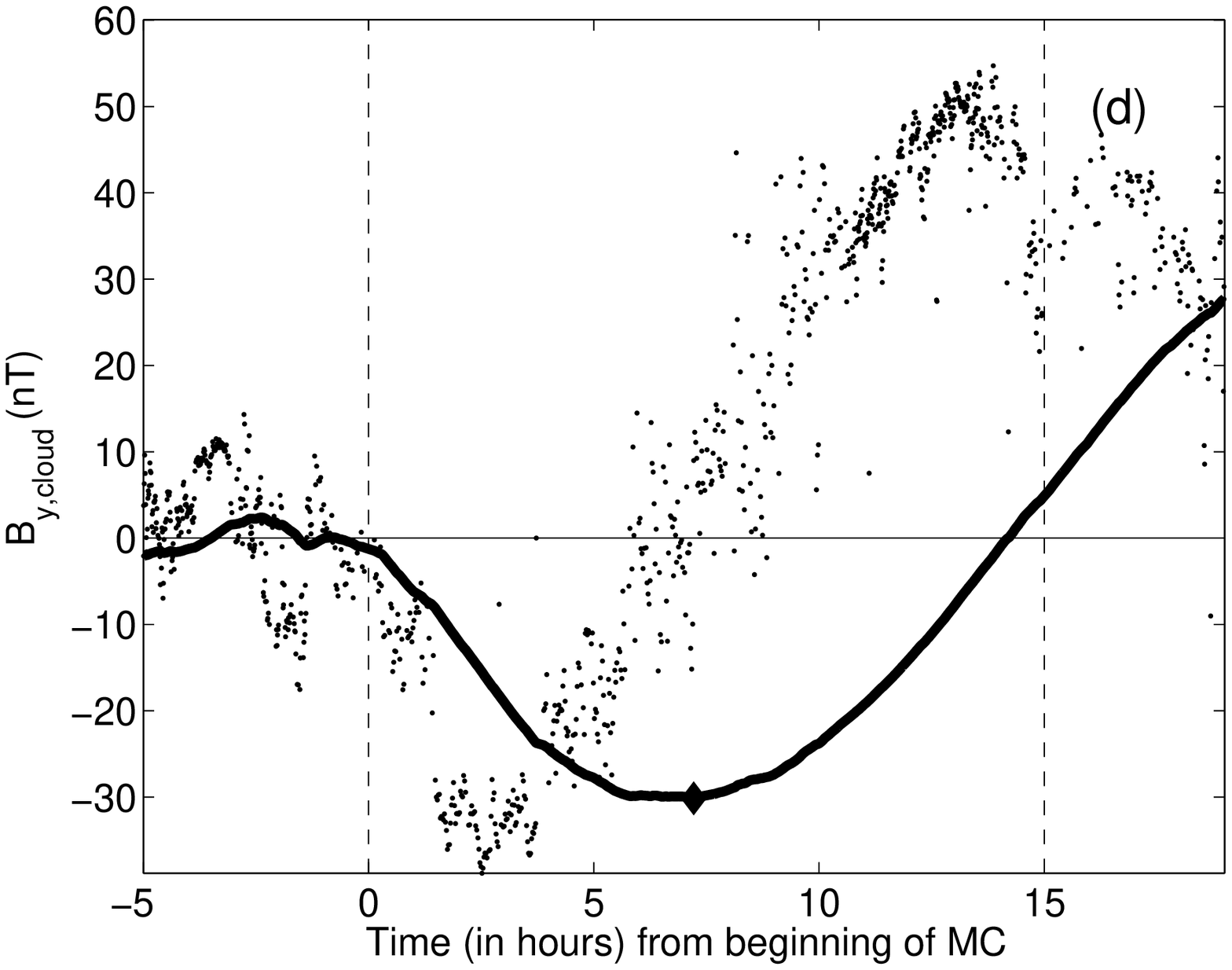}}
\caption{Examples of two analyzed MCs that are not significantly
perturbed by a fast flow. The MC center was observed at 07-Jan-1975
10:39 and 04-Mar-1975 21:37 UT, for panels (a,c) and (b,d),
respectively.   The vertical dashed lines define the MC boundaries.
(a,b) $V_{x}$ is the observed velocity component in the radial
direction from the Sun, expressed in km per second. The straight
line is the linear least square fit of the velocity in the time
interval where an almost linear trend is present (where the
observations are presented as a solid line). The linear fitting is
extrapolated to the borders of the MC, which are marked with
circles. (c,d) $B_{y}$ is the magnetic field component, in nT, both
orthogonal to the MC axis and to the spacecraft trajectory,  while
the solid line represents $F_{y}$, which is the accumulated flux of
$B_{y}$ (Eq.~\ref{Byaccumul}). The extremum of $F_y$ (proxy of the
cloud center) is indicated with diamonds
 (A color version is available in the electronic version).}
 \label{Fig_not_overtaken}
\end{figure*}

MCs interact with their environment during their journey in the
solar wind (SW) from the Sun to the outer heliosphere and, since the
SW pressure (magnetic plus plasma)
 decreases for increasing heliocentric distance, an expansion of MCs is expected.
In a heliospheric frame, the \textit{in situ} observed bulk plasma
velocity typically decreases in magnitude from the front to the back
inside MCs, confirming the expectation that MCs are expanding
objects in the SW. Furthermore, from observations of large samples
of MCs observed at different heliocentric distance, it has been
shown that the size of MCs increases for larger heliocentric
distances \citep[][ and references therein]{Leitner07}.

These structures have an initial expansion from their origin in the
Sun, as shown from observations of radial expansion at the corona;
e.g., an example of the leading edge of a CME traveling faster than
its core is shown in Figure 6 of \cite{Tripathi}. However their
subsequent expansion mainly will be given by the environmental (SW)
conditions as a consequence of force balance \citep{Demoulin09}.

Dynamical models have been used to describe clouds in expansion,
either considering only a radial expansion
\citep[e.g.,][]{Farrugia93,Osherovich93,Farrugia97,Nakwacki08}, or
expansion in both the radial and axial directions
\citep[e.g.,][]{Shimazu02,Berdichevsky03,Dasso07,Nakwacki08b,Demoulin09}.
The main aim of these models is to take into account the evolution
of the magnetic field as the MC crosses the spacecraft. Another goal
is to correct the effect of mixing spatial-variation/time-evolution
in the one-point observations to obtain a better determination of
the MC field configuration. The expansion of several magnetic clouds
has been analyzed previously by fitting different velocity models to
the data
\citep{Farrugia93,Shimazu02,Berdichevsky03,Vandas05,Yurchyshyn06,Dasso07,Mandrini07,Demoulin08}.

The expansion of some MCs is not always well marked with {\it in
situ} velocity measurements. This is in particular the case for
small MCs or those overtaken by fast streams. Slow magnetic clouds,
with velocities lower than or of the order of $400$ km/s, in general
have small sizes, low magnetic field strengths, and only a few of
them present shocks or sheaths \citep[e.g.,][]{Tsurutani04}. Fast
streams overtaking magnetic clouds from behind can compress the
magnetic field in the rear for the overtaken MC, for instance in
some cases forming large structures called merged interaction
regions \citep[e.g.,][]{Burlaga03}. The interaction between a stream
and an MC can affect the internal structure of the cloud
\citep[e.g., as shown from numerical simulations by ][]{Xiong07}.
The difference between the velocities of the front and back
boundaries, called $\Delta V_{\rm obs}$, was frequently used to
qualify how important the expansion of an observed MC is. A larger
$\Delta V_{\rm obs}$ is favorable for the presence of shocks
surrounding the MC, especially for the presence of a backward shock
\citep{Gosling94}. A large $\Delta V_{\rm obs}$ is less important
for the presence of a frontal shock since a frontal shock is also
created by a large difference between the MC global velocity and the
overtaken SW velocity.

The quantity $\Delta V_{\rm obs}$ is a good proxy of the time
variation of the global size of the MC, however, $\Delta V_{\rm
obs}$ does not express how fast the expansion of an element of fluid
is, since $\Delta V_{\rm obs}$ depends strongly on how big the
studied MC is. For example the MC observed by ACE at 1~AU on 29
October 2004 \citep{Mandrini07} is formed by a flux rope with a
large radius, $R\approx 0.17$~AU, and it also has a large $\Delta
V_{\rm obs} \approx 400$~km~s$^{-1}$, and so, at first sight, it can
be qualified as a very rapidly expanding MC. However, let suppose
that the same MC would have most its flux having been reconnected
with the encountered SW during the transit from the Sun, as has been
observed in some cases \citep[e.g.,][]{Dasso06,Dasso07}, so that
only the flux rope core would have been observed as a MC. If the
remaining flux rope would have a radius of only $10^{-2}$~AU, it
would have shown $\Delta V_{\rm obs} \approx 400/17 \approx
24$~km~s$^{-1}$, so it would have been qualified as a slowly
expanding MC.

More generally, small flux ropes are expected to have intrinsically
small $\Delta V_{\rm obs}$, an expectation confirmed by the data
(Fig.~\ref{Fig_correlation}a,b). MCs have a broad range of sizes,
with flux rope radii of a few $0.1$~AU down to a few $10^{-3}$~AU
\citep{Lynch03,Feng07}, and it is necessary to quantify their
expansion rate independently of their size. In this study, we
analyze the expansion of MCs in the inner heliosphere, and find a
non-dimensional expansion coefficient ($\zeta$), which can be
quantified from one-point {\it in situ} observations of the bulk
velocity time profile of the cloud. We demonstrate that $\zeta$
characterizes the expansion rate of the MC, independently of its
size.

We first describe the data used, and then the method to define the
main properties of the MC (Sect.~\ref{Method}).
In Sect.~\ref{Expansion}, we analyze the properties of the MC expansion,
defining a proper expansion coefficient.
 We derive specific properties of two groups of MCs, defined from their
interaction with the SW environment. Then, we relate  the MC
expansion rate to the decrease of the total SW pressure with solar
distance. We summarize our results in Sect.~\ref{summary} and
conclude in Sect.~\ref{conclusion}.

\begin{figure*}[t!]
\centerline{
\includegraphics[width=0.4\textwidth, clip=]{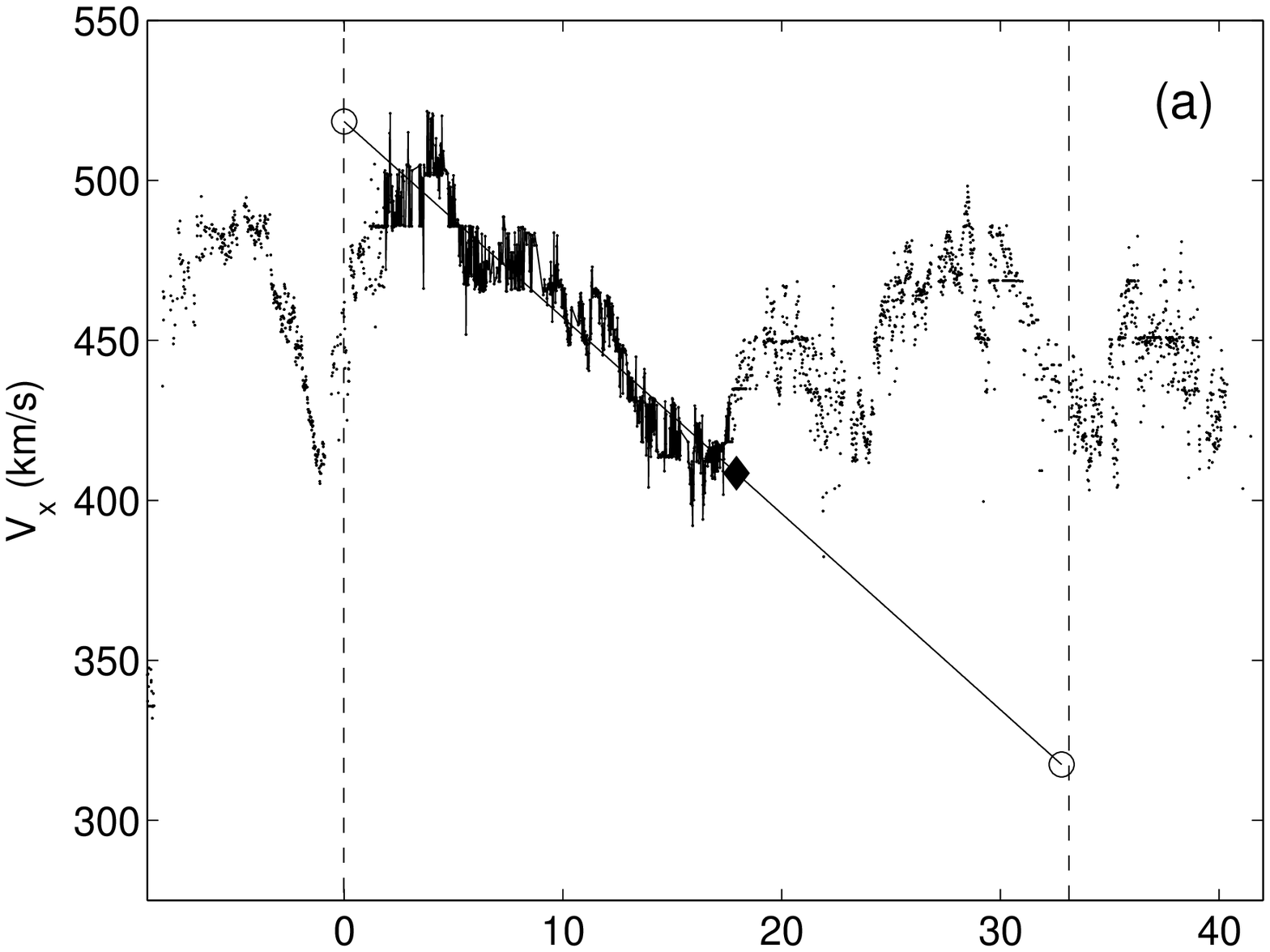}
\includegraphics[width=0.4\textwidth, clip=]{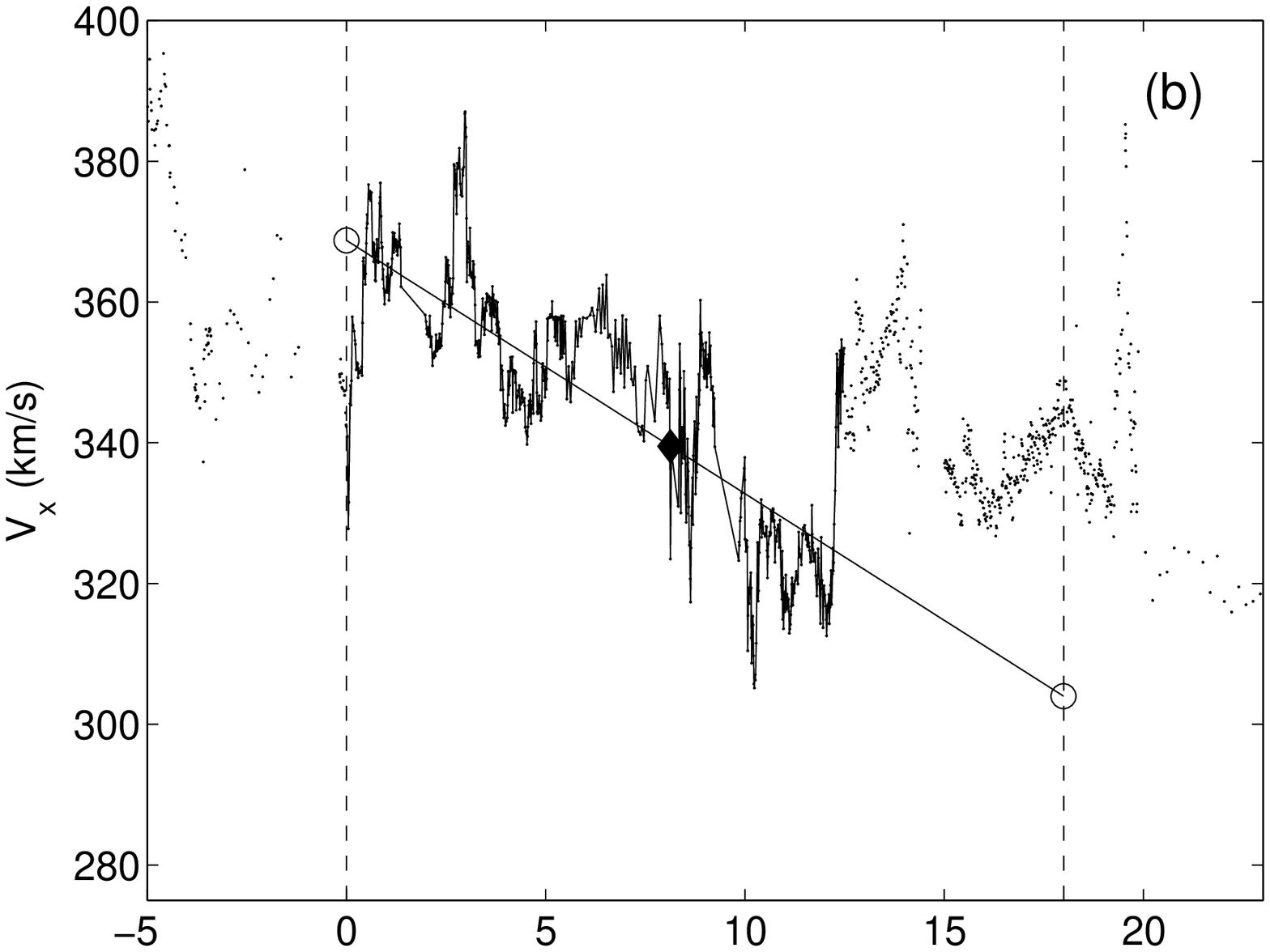}}
\centerline{\includegraphics[width=0.4\textwidth,clip=]{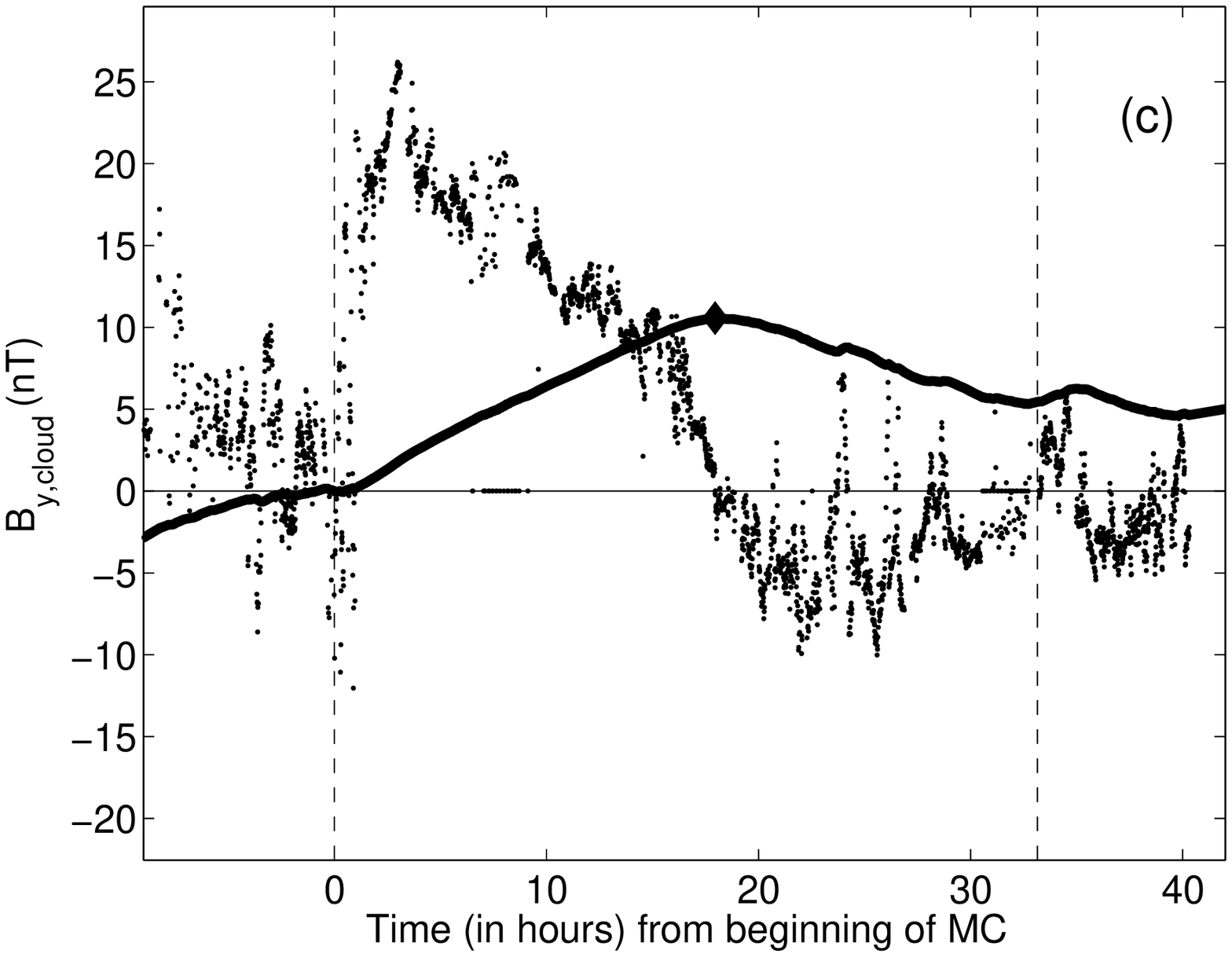}
\includegraphics[width=0.4\textwidth, clip=]{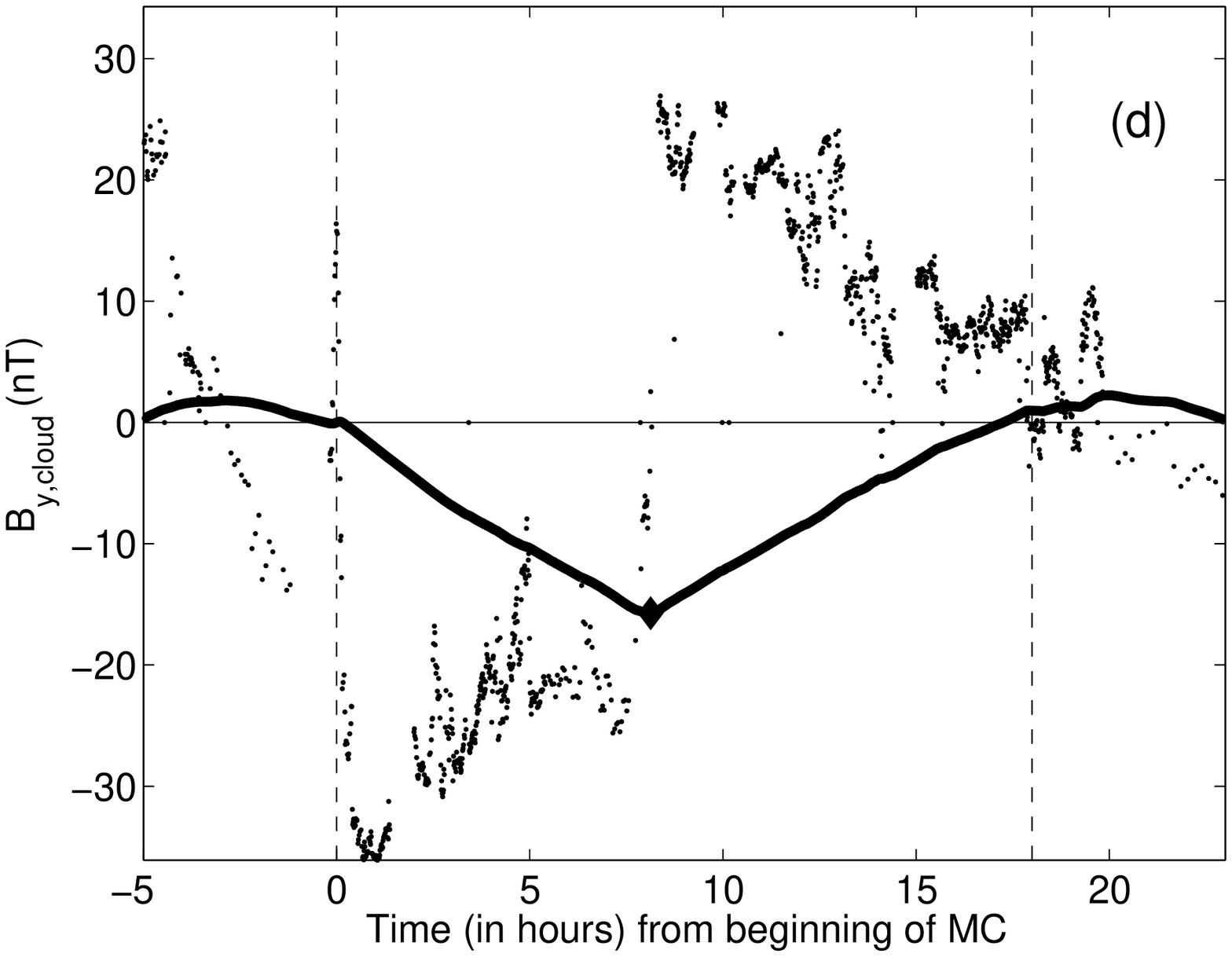}}
\caption{Examples of two analyzed MCs that are perturbed by a fast
flow as seen in the upper panels. The MC center was observed at
30-Jan-1977 03:18 and 23-Jun-1980 12:25 UT, for panels (a,c) and
(b,d), respectively. The external perturbations enter in a
significant part of the MCs. The same quantities, as in
Fig.~\ref{Fig_not_overtaken}, are shown (A color version is
available in the electronic version).}
 \label{Fig_overtaken}

\end{figure*}

\section{Data and method} 
\label{Method}

\subsection{Helios data base} 
\label{M-Helios}
We have studied the MCs reported by different authors from the
Helios 1 and 2 missions \citep{Bothmer98,Liu05,Leitner07}; from
November 1974 to 1985 for Helios 1 and from January 1976 to 1980 for
Helios 2. We analyzed observations of plasma properties
\citep{Rosenbauer77}, in particular bulk velocity and density of
protons \citep{Marsch82}, and magnetic field vector
\citep{Neubauer77}, for a time series with a temporal cadence of $40
\pm 1$ seconds.

  The magnetic and velocity fields observations are in a right-handed
system of coordinates ($\hat{x}$, $\hat{y}$, $\hat{z}$). $\hat{x}$
corresponds to the Sun-Spacecraft direction, $\hat{y}$ is on the
ecliptic plane and points from East to West (in the same direction
as the planetary motion), and $\hat{z}$ points to the North
(perpendicular to the ecliptic plane and closing the right-handed
system).

\subsection{Definition of the MC local frame} 
\label{M-MC-local-frame}
To facilitate the understanding of MC properties, we define a system
of coordinates linked to the cloud in which $\hat{z}_{\rm cloud}$ is
along the cloud axis (with $B_{z, \rm cloud}>0$ at the MC axis).
Since the velocity of an MC is nearly in the Sun-spacecraft
direction and as its speed is much higher than the spacecraft speed
(which can be supposed to be at rest during the cloud observing
time), we assume a rectilinear spacecraft trajectory in the cloud
frame.  The trajectory defines a direction $\hat{d}$ (pointing
toward the Sun); then, we define $\hat{y}_{\rm cloud}$ in the
direction $\hat{z}_{\rm cloud} \times \hat{d}$ and $\hat{x}_{\rm
cloud}$ completes the right-handed orthonormal base ($\hat{x}_{\rm
cloud},\hat{y}_{\rm cloud},\hat{z}_{\rm cloud}$). We also define the
impact parameter, $p$, as the minimum distance from the spacecraft
to the cloud axis.

The observed magnetic field in an MC can be expressed in this local
frame, transforming the observed components ($B_{x}$, $B_{y}$,
$B_{z}$) with a rotation matrix to ($B_{x, \rm cloud}$, $B_{y, \rm
cloud}$, $B_{z, \rm cloud}$). In particular, for $p=0$ and an MC
described by a cylindrical magnetic configuration, i.e. $\vec{B}(r)
= B_z(r) \hat{z} + B_\phi(r) \hat{\phi}$, we have $\hat{x}_{\rm
cloud} = \hat{r}$ and $\hat{y}_{\rm cloud} = \hat{\phi}$ when the
spacecraft leaves the cloud.  In this particular case, the magnetic
field data will show: $B_{x, \rm cloud}=0$, a large and coherent
variation of $B_{y, \rm cloud}$ (with a change of sign), and an
intermediate and coherent variation of $B_{z, \rm cloud}$, from low
values at one cloud edge, with the largest value at its axis and
returning to low values at the other edge.

More generally, the local system of coordinates is especially useful
when $p$ is small compared to the MC radius ($R$) since the
direction of the MC axis can be found using a fitting method or
applying the minimum variance (MV) technique to the normalized time
series of the observed magnetic field \citep[e.g.][ and references
therein]{Dasso06}. In particular, from the analysis of a set of
cylindrical synthetic MCs, \citet{Gulisano07} found that the
normalized MV technique provides a deviation of the real orientation
of the main MC axis of less than $10 ^\circ$ even for $p$ as large
as $50\%$ of the MC radius.

\subsection{Definition of the MC boundaries} 
\label{M-MC-boundaries}
As the first step of an iterative process, we choose the MC boundaries reported in the literature, and perform a normalized minimum variance analysis to find the local frame of the MC.  We then analyze the magnetic field components in the local frame,
and redefine the boundaries of each MC, according to the expected typical behavior of the axial field ($B_{z,cloud}$, having its maximum near the center and decreasing toward the MC boundaries), the azimuthal field ($B_{y,cloud}$, maximum at one of the borders, minimum at the other one and changing its signs near the cloud center),
and $B_{x,cloud}$, which is expected to be small and with small variations \citep[see][]{Gulisano07}.
 Moreover, in the MC frame it is easier to differentiate the SW and MC sheath magnetic field, which is fluctuating, from the MC field, which has a coherent and expected behaviour for the three field components.
We then moved the borders in order to reach these properties of the local field components, and performed the same procedure iteratively to find an improved
orientation, using several times the minimum variance technique.
We applied this procedure to each MC of our sample.

\begin{table}
\caption{List of MC events.}

 \label{table_res}
\begin{center}
\begin{tabular}{cc@{ }c|r@{.}lr@{.}l} 
 \hline
S/C & $T_c$ & Group & \multicolumn{2}{c}{$\Delta V_x/(t_{out}-t_{in})$}
                    & \multicolumn{2}{c}{$\zeta$} \\
    & d-m-y h:m (UT) & & \multicolumn{2}{c}{km~s$^{-1}$~h$^{-1}$}
                      & \multicolumn{2}{c}{} \\

 \hline
    H1 & 07-{\tt Jan}-1975 10:39 & N  &~~~~ 11&1 & ~~1&0 \\
    H1 & 04-{\tt Mar}-1975 21:37 & N  &  8&75    &  0&63\\
    H1 & 02-{\tt Apr}-1975 09:00 & P  & -6&84    & -1&1\\
    H1 & 05-{\tt Jul}-1976 14:20 & P  &  0&15    &  0&04\\
    H1 & 30-{\tt Jan}-1977 03:18 & P  &  6&13    &  1&4\\
    H1 & 31-{\tt Jan}-1977 00:26 & P  &  3&68    &  0&75\\
    H1 & 20-{\tt Mar}-1977 01:13 & P  &  4&28    &  0&52\\
    H1 & 09-{\tt Jun}-1977 01:22 & N  &  4&26    &  0&86\\
    H1 & 09-{\tt Jun}-1977 10:33 & P  &  3&71    &  0&9\\
    H1 & 28-{\tt Aug}-1977 21:36 & P  &  0&92    &  0&25\\
    H1 & 26-{\tt Sep}-1977 03:45 & N  &  14&8    &  0&97\\
    H1 & 01-{\tt Dec}-1977 20:15 & P  & -0&6     & -0&11\\
    H1 & 03-{\tt Jan}-1978 19:21 & N  &  14&1    &  0&83\\
    H1 & 16-{\tt Feb}-1978 07:25 & N  &  7&66    &  1&5\\
    H1 & 02-{\tt Mar}-1978 12:33 & N  &  4&99    &  0&89\\
    H1 & 30-{\tt Dec}-1978 01:50 & N  &  12&4    &  1&1\\
    H1 & 28-{\tt Feb}-1979 10:13 & N  &  5&7     &  0&78\\
    H1 & 03-{\tt Mar}-1979 18:49 & N  &  5&51    &  0&58\\
    H1 & 28-{\tt May}-1979 23:07 & p  & -4&57    & -0&37\\
    H1 & 01-{\tt Nov}-1979 09:09 & P  &  2&62    &  0&5\\
    H1 & 22-{\tt Mar}-1980 21:25 & P  &  7&78    &  2&0\\
    H1 & 10-{\tt Jun}-1980 20:31 & P  & -6&27    & -0&62\\
    H1 & 20-{\tt Jun}-1980 05:47 & P  &  4&24    &  0&48\\
    H1 & 23-{\tt Jun}-1980 12:25 & P  &  3&6     &  0&75\\
    H1 & 27-{\tt Apr}-1981 11:55 & N  &  11&9    &  1&3\\
    H1 & 11-{\tt May}-1981 23:30 & N  &  21&6    &  1&0\\
    H1 & 27-{\tt May}-1981 05:43 & N  &  6&83    &  0&89\\
    H1 & 19-{\tt Jun}-1981 05:05 & N  &  25&5    &  0&75\\
    H2 & 06-{\tt Jan}-1978 06:50 & N  &  6&97    &  0&84\\
    H2 & 30-{\tt Jan}-1978 05:02 & P  &  11&7    &  1&2\\
    H2 & 07-{\tt Feb}-1978 13:45 & N  &  2&42    &  0&68\\
    H2 & 17-{\tt Feb}-1978 02:32 & N  &  3&23    &  0&8\\
    H2 & 24-{\tt Apr}-1978 11:54 & P  &  15&4    &  1&0\\
\hline
    H1 &       & all &  6&2 $\pm$ 7.3  &  0&66 $\pm$ 0.65\\
    H1 &       & P   &  1&3 $\pm$ 4.5  &  0&39 $\pm$ 0.81\\
    H1 &       & N   & 11&1 $\pm$ 6.3  &  0&94 $\pm$ 0.25\\
 \hline
    H2 &       & all &  8&0 $\pm$ 5.6  &  0&90 $\pm$ 0.19\\
    H2 &       & P   & 13&6 $\pm$ 2.6  &  1&09 $\pm$ 0.13\\
    H2 &       & N   &  4&2 $\pm$ 2.4  &  0&77 $\pm$ 0.08\\
 \hline
    Both &     & all &  6&5 $\pm$ 7.   &  0&70 $\pm$ 0.61\\
    Both &     & P   &  2&9 $\pm$ 6.   &  0&48 $\pm$ 0.79\\
    Both &     & N   & 10&0 $\pm$ 6.4  &  0&91 $\pm$ 0.23\\
 \hline
\end{tabular}
\end{center}
The first column in Table 1 indicates the spacecraft (Helios 1 or
Helios 2), $T_c$ in the second column is the time for the
observation of the MC center (or for the minimum approach distance).
The MCs are separated in two groups: perturbed (P) and non-perturbed
(N) by a fast SW stream. $\Delta V_x/(t_{out}-t_{in})$ is minus the
fitted slope of the temporal velocity profile (see, Eq.~\ref{dV})
and $\zeta$ is the non-dimensional expansion coefficient
(Eq.~\ref{zeta}). $\Delta V_x/(t_{out}-t_{in})<0$ means observed
compression of the MC. The average values and the standard
deviations are given at the bottom.
 \end{table}

\subsection{Characterization of the MC expansion} 
\label{M-MC-expansion}

    Most MCs have a higher velocity in their front than in their back,
showing that they are expanding magnetic structures in the SW.
About half of the studied MCs have well defined linear profile $V_{x}(t)$
(Fig.~\ref{Fig_not_overtaken}a,b), while for the other half, $V_{x}(t)$
is nearly linear only in a part of the MC which includes
the MC center (Fig.~\ref{Fig_overtaken}a,b).
The distortions of $V_{x}(t)$ are more frequently due to an overtaking
faster SW flow in the back of the MC.

 We split the data set in two groups:
non-perturbed MCs for cases where the velocity profile presents a
linear trend in more than $75$\% of the full size of the MC and
perturbed MCs for cases where this is not satisfied. There are
almost as many perturbed as non-perturbed
 MCs, considering data from each spacecraft separately and both of them together.
  The measured temporal profile $V_{x}(t)$ is fitted using a least square
fit with a linear function of time,
   \begin{equation}  \label{linear_fit}
   V_{x, \rm fit}(t) = V_{\rm o, fit} + \dVxdt ~t \,,
   \end{equation}
where $\dVxdt$ is the fitted slope of the linear function.
We always keep the fitting range inside the MC,
but restrict it to the most linear part of the observed profile.
This choice minimizes the effect of the interacting flows
(but it does not fully remove it, since a long term interaction can,
a priori, change the expansion rate of the full MC).

The linear fit is used to define the velocities $V_{x, \rm
fit}(t_{in})$ and $V_{x, \rm  fit}(t_{out})$ at the MC boundaries
(Sect.~\ref{M-MC-boundaries}). Then, we define the full expansion
velocity of an MC as:
    \begin{equation}  \label{dV}
    \Delta V_x = V_{x, \rm  fit}(t_{in}) - V_{x, \rm  fit}(t_{out}) \,.
    \end{equation}
For non-perturbed MCs, $\Delta V_x$, is very close to the observed
velocity difference  $V_x(t_{in}) - V_x(t_{out})$, see e.g. upper
panels of Fig.~\ref{Fig_not_overtaken}. For perturbed MCs, this
procedure minimizes the effects of the perturbations entering the
MC. Then, the expansion velocity is defined consistently for the
full set of MCs.

\begin{figure*}[t!]
\centerline{
\includegraphics[width=0.4\textwidth, clip=]{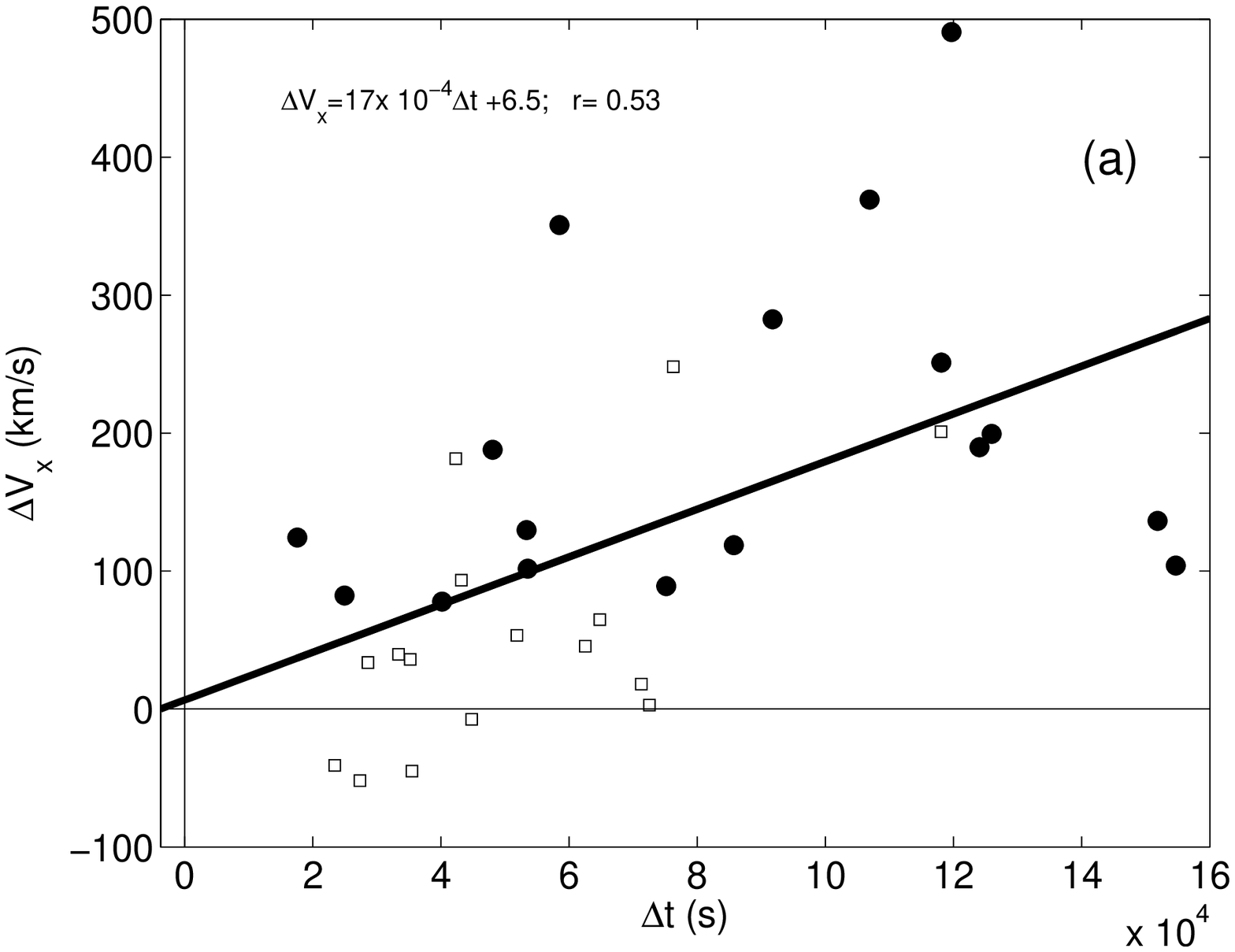}
\includegraphics[width=0.4\textwidth, clip=]{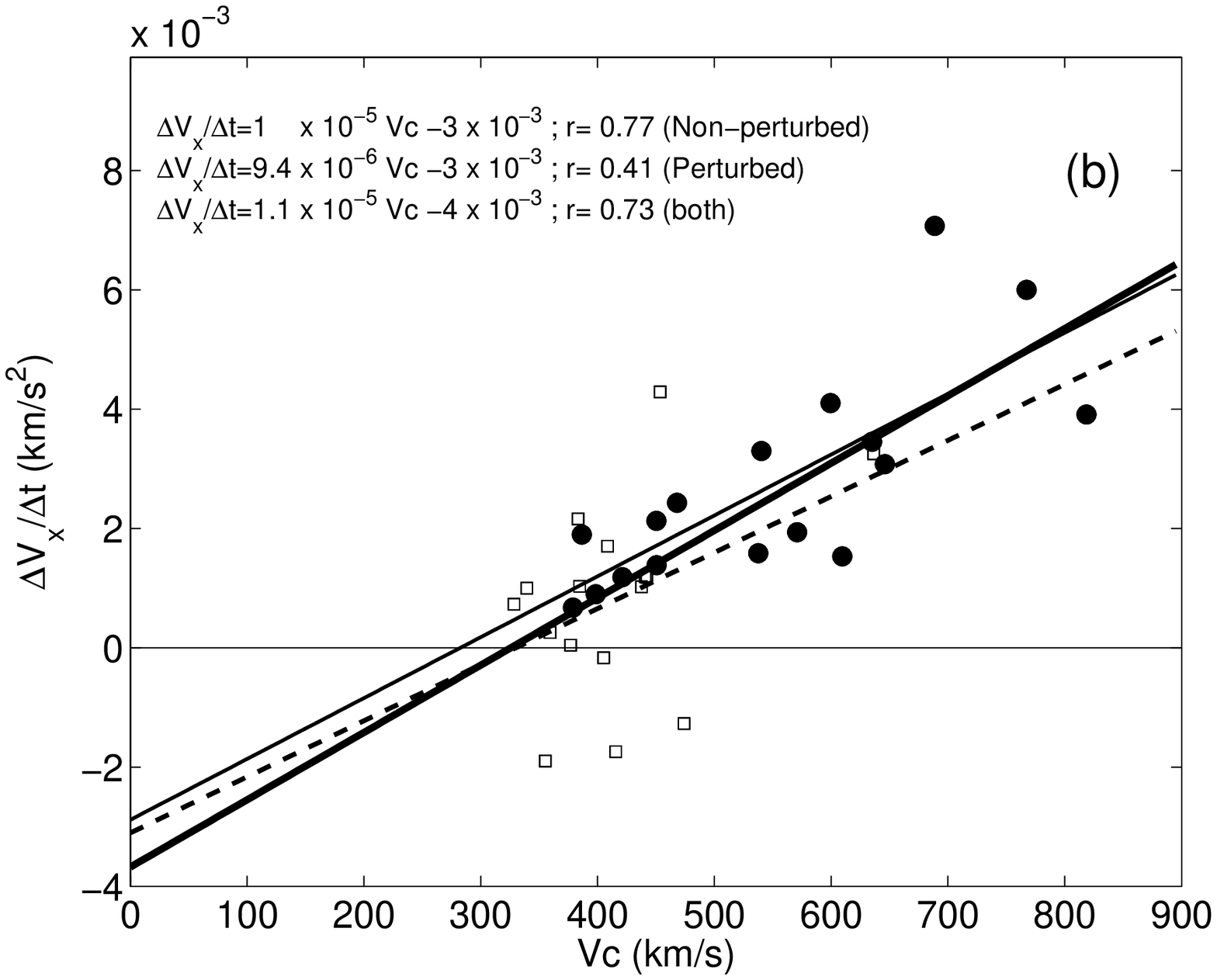}}
\centerline{\includegraphics[width=0.4\textwidth,clip=]{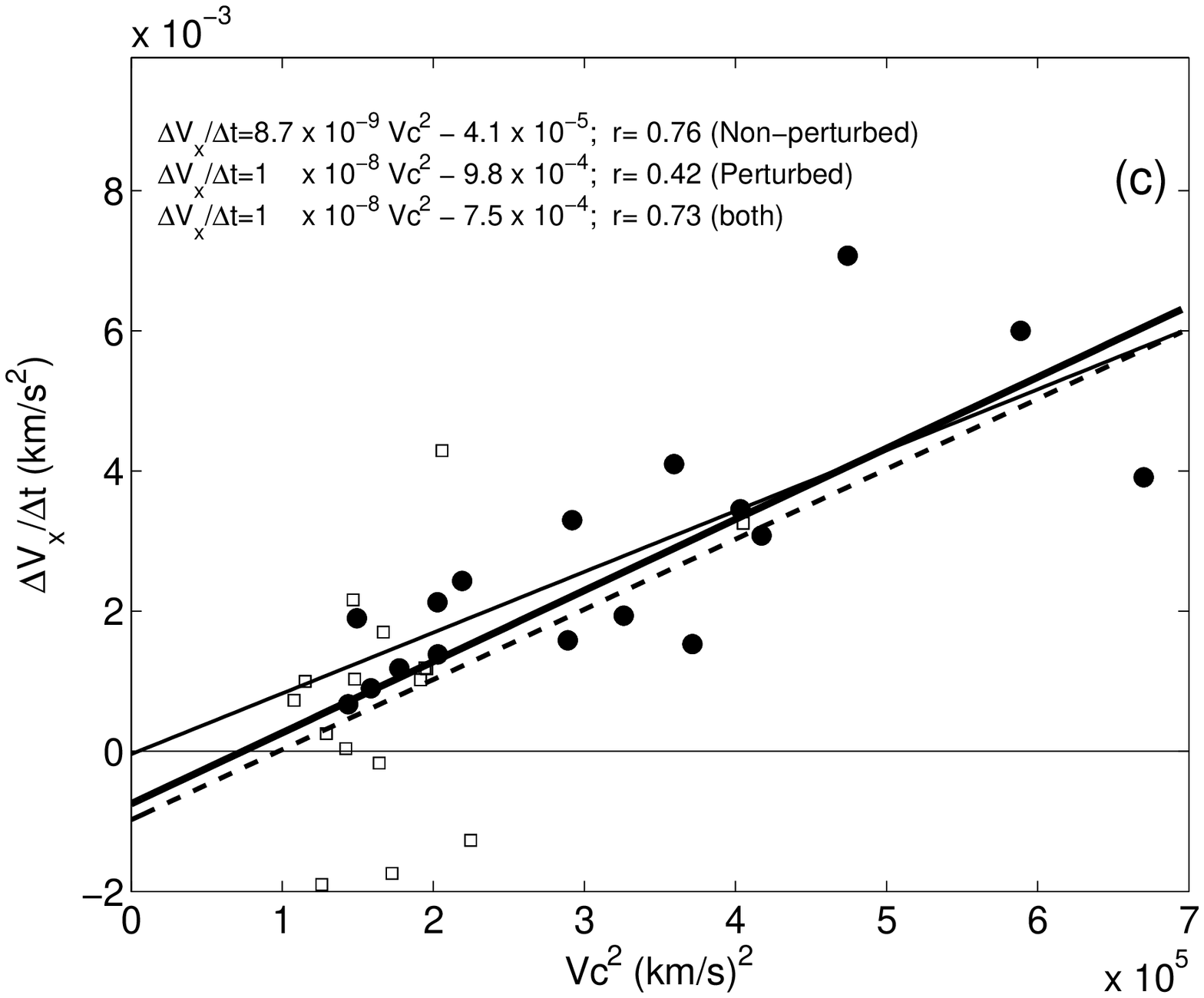}
\includegraphics[width=0.4\textwidth, clip=]{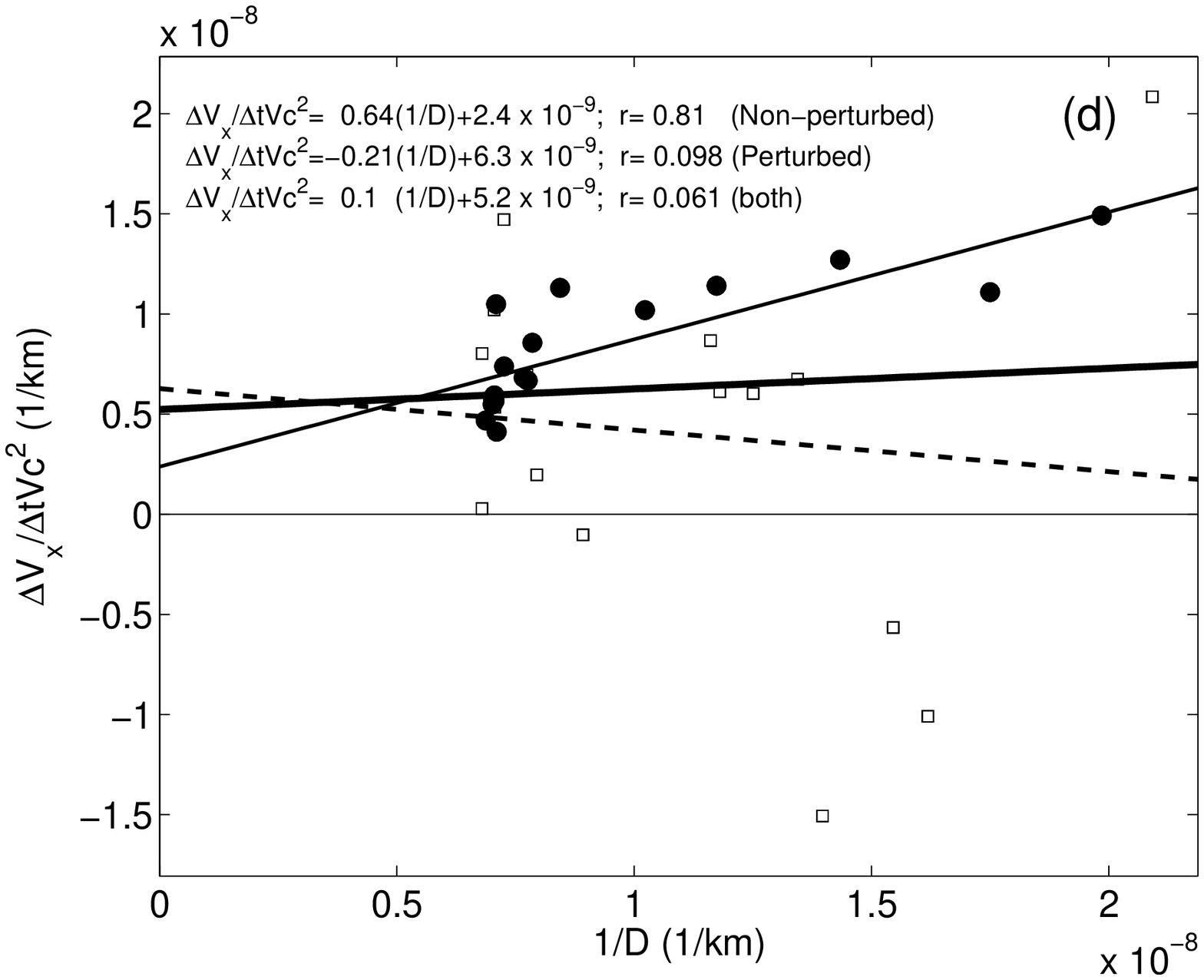}}
\caption{ The panels a-d show the correlation analysis between
proxies for MC expansion with different physical quantities.  The
MCs are separated in two groups: perturbed (empty square symbol) and
non-perturbed (filled circle symbol). The straight lines are the
result of a least square fit for perturbed (dashed line),
non-perturbed (thin continuous line), and for both set of MCs in the
list of events shown in Table~\ref{table_res} (thick continuous
line). $\Delta V_x $ is defined by Eq.~(\ref{dV}), $\Delta t =t_{\rm
out}- t_{\rm in}$, $V_c$ is the velocity of the MC center (or at the
closest approach distance), and $D$ is the distance to the Sun.  The
fitted values and the obtained correlation coefficient ($r$) are
included as insets, considering the different groups. More
significant correlation is present for the non-perturbed cases
 (A color version is available in the electronic version).
 }
 \label{Fig_correlation}
\end{figure*}

\subsection{MC size} 
\label{M-MC-size}
  From the determination of the boundaries described above,
we can estimate the size, $S$, of the flux rope along $\hat{x}$ (the Sun-Spacecraft direction). $S$ is computed as the time duration of observation of the MC multiplied by the MC velocity at the center of the structure (see end of section ~\ref{M-MC-center}).

We perform a linear least square fit in log-log plots of $S$ as a
function of the distance to the Sun ($D$). As expected, the size has
a clear dependence with $D$:
\begin{eqnarray}
S_{\rm full\ set}       & = & (0.23 \pm 0.01) ~D^{0.78 \pm 0.12}     \nonumber    \\
S_{\rm non-perturbed}  & = & (0.32 \pm 0.02) ~D^{0.89 \pm 0.15}     \label{S(D)} \\
S_{\rm  perturbed}      & = & (0.16\pm 0.01) ~D^{0.45 \pm 0.16} \,,
\nonumber
\end{eqnarray}
where $S$ and $D$ are in AU.
Our results are compatible within the error bars with previous studies:
\begin{eqnarray}
d_{\rm Bothmer}    & = & (0.24 \pm 0.01) ~D^{0.78 \pm 0.10}      \nonumber \\
S_{\rm Liu}        & = & (0.25 \pm 0.01) ~D^{0.92 \pm 0.07}      \label{Sothers(D)} \\
S_{\rm Wang}       & = & ~0.19 \qquad \quad\,\, ~D^{0.61}            \nonumber \\
d_{\rm Leitner}     & = & (0.20 \pm 0.02) ~D^{0.61 \pm 0.09} \,,  \nonumber
\end{eqnarray}

where $d$ is an estimation of the true diameter of the MC. Our
results are closer to \citet{Bothmer98} and \citet{Liu05} who
analyzed MCs and ICMEs, respectively. The larger difference is
between our results and the last two ones while they are based on
the most different sets, as follows. \citet{Wang05b} studied a large
set of ICMEs defined only by a measured temperature lower by a
factor of 2 than expected in the SW with the same speed
\citep[e.g.,][]{Richardson04}. This set included MCs, but it is
dominated by non-MC ICMEs. Conversely, \citet{Leitner07} analyzed
only MCs, with a strict classical definition. They fitted the
magnetic field observations with a classical cylindrical linear
force-free field, then they found the impact parameter and the
orientation of the MC axis to estimate the true diameter, $d$, of
the MCs. So the selected events and the method of
\citeauthor{Wang05b} and \citeauthor{Leitner07} are noticeably
different.  We also note that \citet{Leitner07} found a larger
exponent, $1.14 \pm 0.44$, when they restrict their data to $D \leq
1$~AU. This indicates that the relation is not strictly a power-law
and this could be the main origin of the different results (which
are so dependent on the distribution of events with solar distance
in the selected sets).

\begin{figure*}[t!]
\centerline{
\includegraphics[width=0.4\textwidth, clip=]{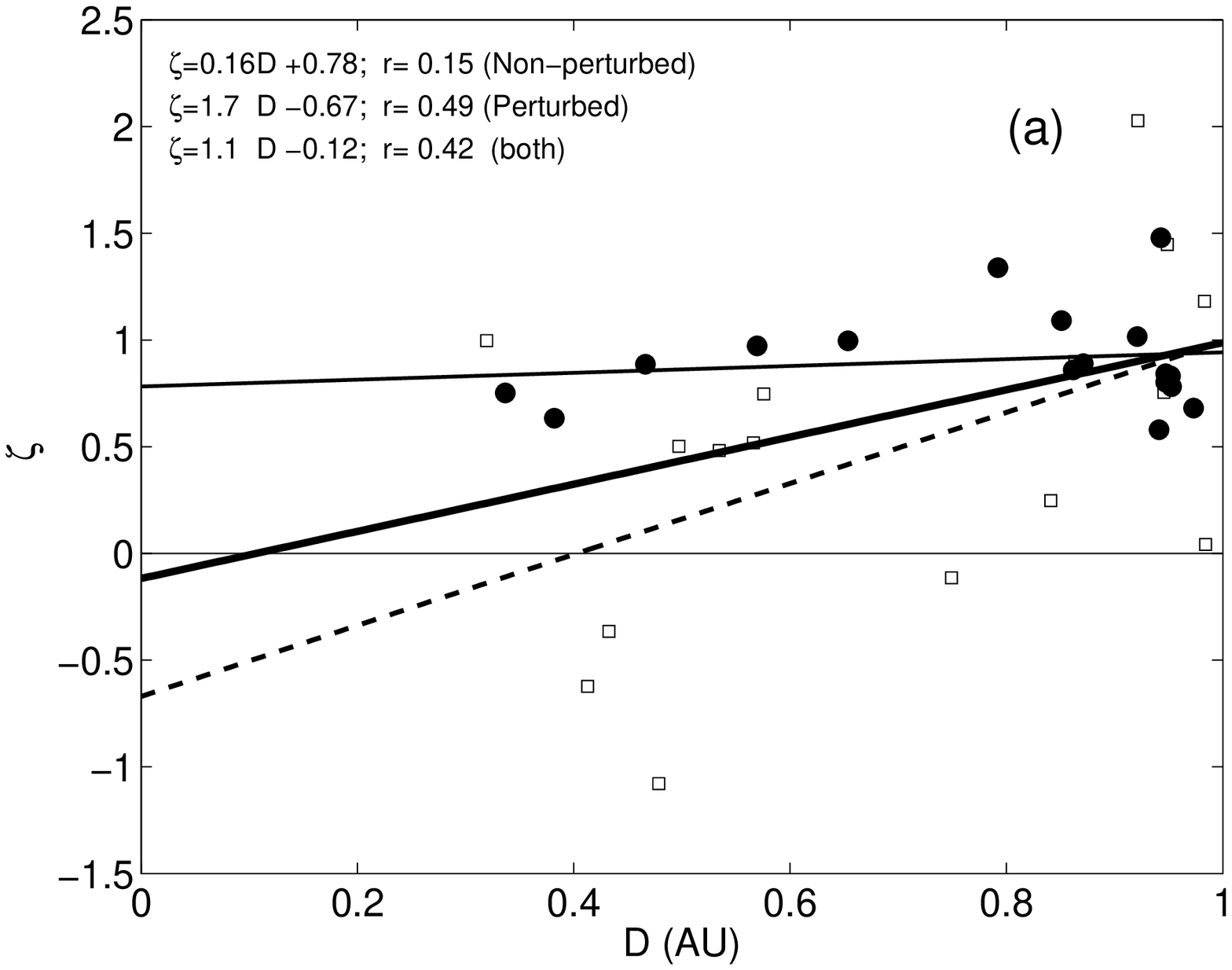}
\includegraphics[width=0.4\textwidth, clip=]{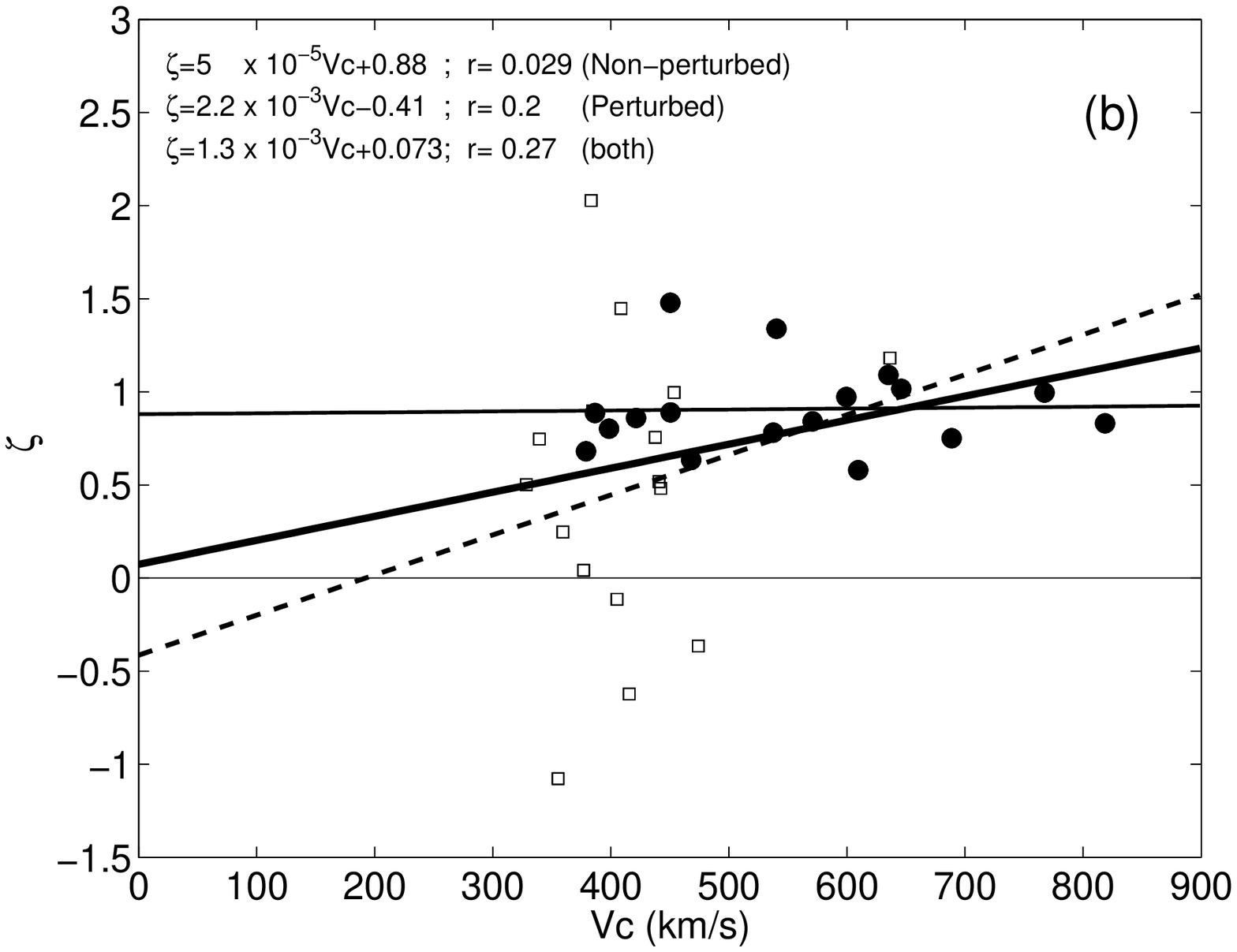}}
\centerline{\includegraphics[width=0.4\textwidth,clip=]{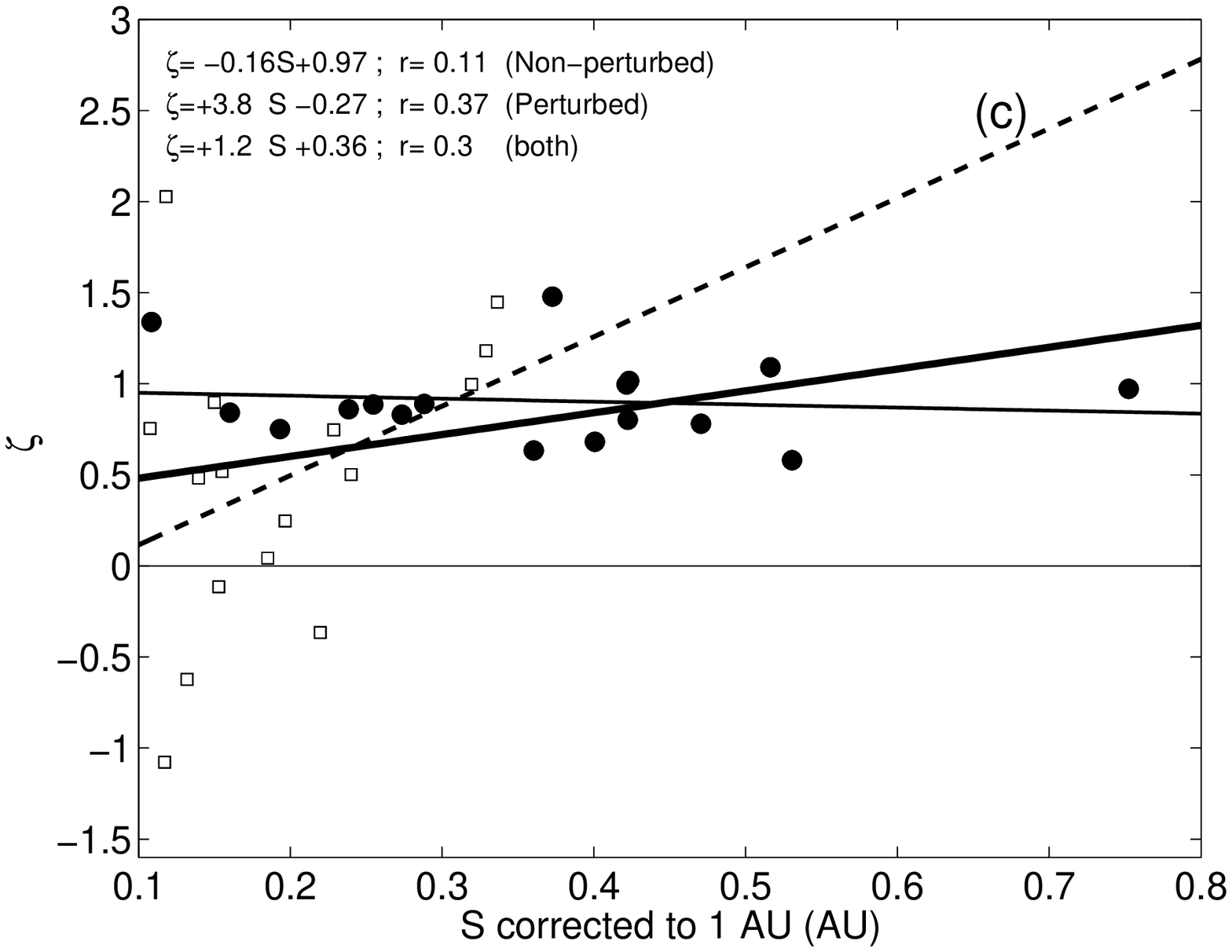}
\includegraphics[width=0.4\textwidth, clip=]{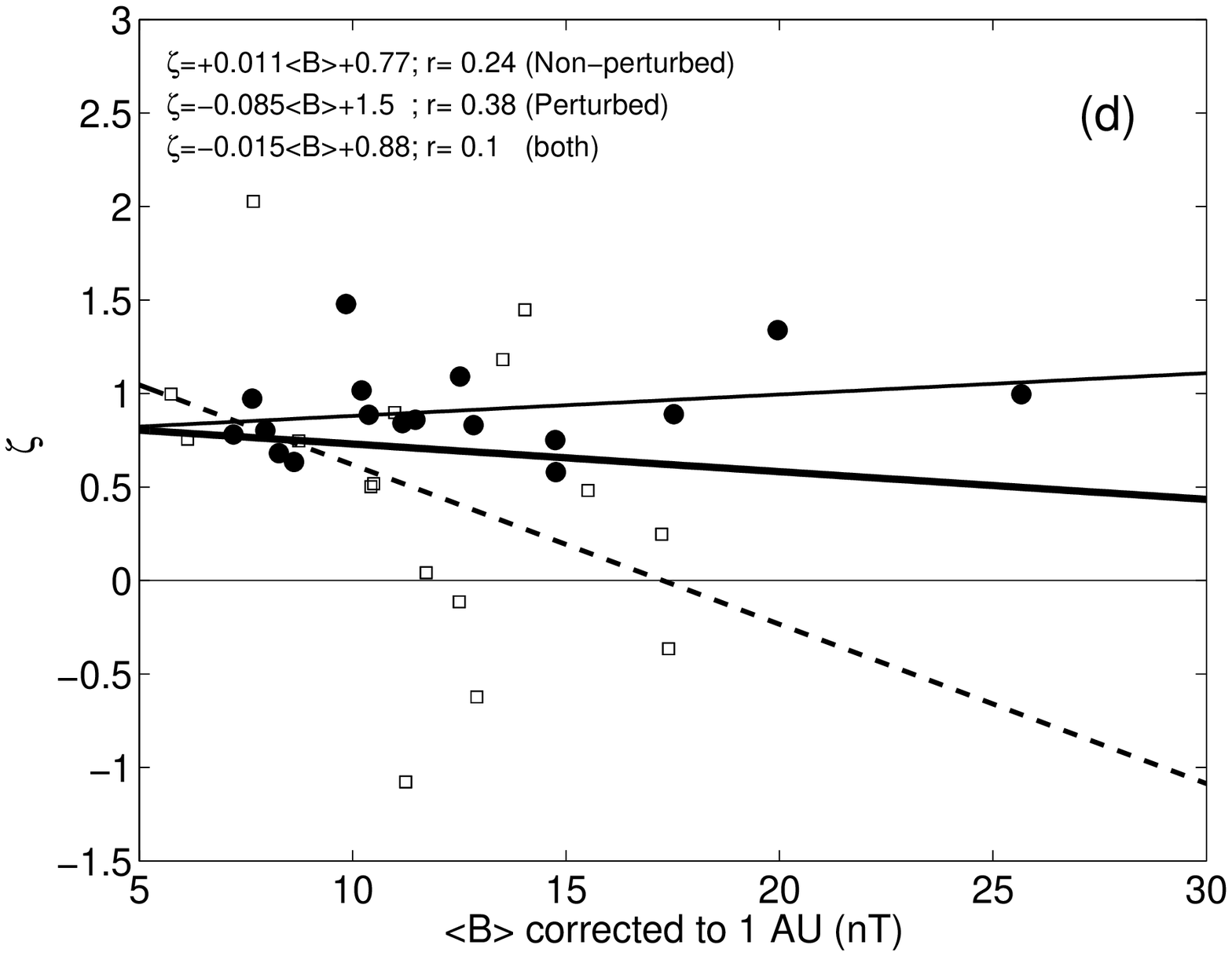}}
  \caption{Panels a-d show the correlation analysis that tests for the dependence of the non-dimensional
   expansion factor $\zeta$ (Eq.~\ref{zeta}) as a function of other parameters.
Perturbed and non-perturbed MCs are represented as in
Fig.~\ref{Fig_correlation}. $S_{\rm corrected~to~1~AU}$ and
$<B>_{\rm corrected~to~1~AU}$ are normalized to 1 AU using the size
and field strength dependence on the distance, according to the
relationship given in Eq.~(\ref{S+Bcorrected}).
 (A color version is available in the electronic version).
 }
 \label{Fig_zeta}
\end{figure*}

\subsection{Magnetic field strength} 
\label{M-MC-<B>}
Another important characteristic of MCs is their magnetic field strength.
We define the average field $<B>$ within each MC and proceed as above with $S$.
$<B>$ has an even stronger dependence upon distance:
\begin{eqnarray}
<B>_{\rm  full\ set}      & = & (10.9\, \pm\ 0.4)\quad   D^{-1.85 \pm 0.07}     \nonumber  \\
<B>_{\rm  non-perturbed} & = & (11.4\, \pm\ 0.5)\quad   D^{-1.85 \pm 0.11}     \label{B(D)} \\
<B>_{\rm  perturbed}     & = & (10.4\, \pm\ 0.6)\quad D^{-1.89 \pm
0.10} \,, \nonumber
\end{eqnarray}
where the units of $<B>$ and $D$ are nT and AU respectively.
Our results have a stronger dependence on $D$ than previous results:
\begin{eqnarray}
<B>_{\rm Liu}             & = & (7.4 \pm 0.4)       ~D^{-1.40 \pm 0.08}     \nonumber \\
<B>_{\rm Wang}            & = & ~8.3 \qquad \,\,\,\,\,\,   D^{-1.52}              \label{Bothers(D)} \\
<B>_{\rm Leitner}         & = & (19 \pm 1.4)\       ~D^{-1.30 \pm 0.09} \,. \nonumber
\end{eqnarray}
Again the strongest difference exists between our results and the
ones of \citet{Leitner07}. The origin of this difference is expected
to be the same as for the size. Indeed for $D \leq 1$~AU,
\citet{Leitner07} found a more negative exponent, $-1.64 \pm 0.4$,
closer to our results, as ocurred for the size.

    The typical expansion speed in MCs is of the order of half the
Alfv\'en speed \citep[e.g.,][]{Klein82}. In our studied set of MCs,
we have also found that the expansion speed was lower than the
Alfv\'en speed [not shown]. This is a necessary condition to expect
that the magnetic field evolves globally, adapting its initial
magnetic field during its expansion, because the Alfv\'en speed is
the velocity of propagation of information through the magnetic
structure.

\subsection{MC center and translation velocity} 
\label{M-MC-center}

Following \citet{Dasso06}, we define the accumulative flux per unit length $L$
(along the MC axial direction):
    \begin{equation} \label{Byaccumul}
    \frac{F_y(t_1,t_2)}{L}
    = \int_{t_1}^{t_2} B_{y,cloud}(t') ~V_{x,cloud}(t') ~\rmd t' \,.
    \end{equation}
Here we neglect the evolution of the magnetic field during the
spacecraft crossing period (so also the ``aging'' effect, see
\citealp{Dasso07}). The set of field lines, passing at the position
of the spacecraft at $t_1$, with the hyphotesis of symmetry of
translation along the main axis, defines a magnetic flux surface,
which is wrapped around the flux rope axis. Then, any magnetic flux
surface will be crossed at least twice by the spacecraft, once at
$t_1$ and once at $t_2$ defined by $F_y(t_1,t_2)=0$.
Then, this property of $F_y(t_1,t_2)$ permits us to associate any
out-bound position, within the flux rope, to its in-bound position
belonging to the same magnetic-flux surface.
The global extremum of $F_y(t_1,t)$, for $t_1$ having a fixed value,
locates the position where the spacecraft trajectory has the closest
approach distance to the MC axis (MC center). This position can also
be found directly from $B_{y,cloud}(t)$, where $B_{y,cloud}$ crosses
zero. Nevertheless, using the integral quantity $F_y$ has the
advantage of decreasing the fluctuations and of outlining the global
extremum, compared to the local extrema (see lower panels of
Figs.~\ref{Fig_not_overtaken} and \ref{Fig_overtaken}).\\
The velocity at the MC center ($V_c$) is computed from the fitted linear
regression  Eq.~(\ref{linear_fit}) evaluated at the time when the spacecraft
reaches the MC center.

\section{Expansion rate of MCs} 
\label{Expansion}

\subsection{Correlation involving the expansion velocity } 
\label{E-Correlation }

From here on, we classify the MCs belonging to the full set of
events, according to the quality of their velocity and magnetic
observations. If they were too noisy or with a lot of data gaps, we
exclude them from the following study, keeping only those MCs with
relatively good quality (listed in Table~\ref{table_res}).

   $\Delta V_x$, as defined by Eq.~(\ref{dV}), characterizes
the expansion speed of the crossed MC. However, as outlined in
Sect.~\ref{Introduction}, $\Delta V_x$ is expected to be strongly
correlated with the MC size, so that it does not express directly
how fast a given parcel of plasma is expanding in the MC. We
therefore define below, after a few steps, a better measure of the
expansion rate.
The size of an MC is proportional to $\Delta t =t_{\rm out}- t_{\rm
in}$ and to $V_c$. Fig.~\ref{Fig_correlation}a shows a clear
positive correlation between $\Delta V_x$ and $\Delta t$. Moreover
the least square fit of a straight line for the full set of MCs
gives a fitted curve passing in the vicinity of the origin (within
the uncertainties present on the slope).  This affine correlation is
then removed by computing $\Delta V_x / \Delta t$. This quantity
also shows, as expected,  a positive correlation with $V_c$
(Fig.~\ref{Fig_correlation}b), but differently above, the fitted
straight line stays far from the origin, so we cannot simply remove
the correlation by dividing $\Delta V_x / \Delta t$ by $V_c$.
However, its dependence on $V_c^2$ brings the fitted straight line
close enough to the origin (within the uncertainties of the fit,
Fig.~\ref{Fig_correlation}c) so that $\Delta V_x / (\Delta t V_c^2)$
is a meaningful quantity. These correlations are present for both
groups of MCs, but they are much stronger for non-perturbed MCs
(Fig.~\ref{Fig_correlation}b,c).
Other correlations have being attempted with the above methodology.
Either there is no significant correlation, or the fitted curve
passes far from the origin. There is still the exception that
$\Delta V_x / (\Delta t V_c^2)$ has an affine correlation with $1/D$
for non-perturbed MCs (Fig.~\ref{Fig_correlation}d).

\subsection{Non-dimensional expansion rate} 
\label{E-zeta}

The above empirical correlation analysis suggests that we can define
the non-dimensional expansion rate as the quantity:
   \begin{equation}   \label{zeta}
   \zeta = \frac{\Delta V_x}{\Delta t} \frac{D}{V_c^2} \,.
   \end{equation}
The first steps were to remove the MC size dependence, while the
last step could be further justified by the need to have a
non-dimensional coefficient. Finally, it is remarkable that the
correlation analysis of the MC data leads to the definition of the
same variable, $\zeta$, as the theoretical analysis of
\citet{Demoulin09}.

 We next verify that $\zeta$ is no longer dependent on $\Delta t $, $V_c$, $D$ or some combination of them.
Figures~\ref{Fig_zeta}a,b show two examples of this exploration.
Indeed, the non-perturbed MCs show almost no correlation, while
there are still some correlations when the perturbed MCs are
considered. Figure~\ref{Fig_zeta}b also shows that even for slow MCs
(see Sect.~\ref{Introduction}), the non-dimensional expansion rate
($\zeta$) is not correlated with $V_c$.

Still, does $\zeta$ depend on the properties of the MC considered?
To test this we first need to remove the distance dependence on $S$
and $<B>$, found in Sect.~\ref{M-MC-size} and~\ref{M-MC-<B>}, by
defining values at a given solar distance (here taken at 1~AU). We
use:
   \begin{eqnarray}
   S_{\rm corrected~to~1~AU}   & = &   S ~ D^{-0.8}      \nonumber   \\
   <B>_{\rm corrected~to~1~AU} & = & <B> ~ D^{+1.8} \,. \label{S+Bcorrected}
   \end{eqnarray}

We find that there is no significant correlation between $\zeta$ and
$S_{\rm corrected~to~1~AU}$, as well as between $\zeta$ and
$<B>_{\rm corrected~to~1~AU}$ for the non-perturbed MCs
(Fig.~\ref{Fig_zeta}c,d). Taking other exponents, in the vicinity
$(\pm 0.4)$ of the exponents used in Eq.~\ref{S+Bcorrected}, we also
find very low correlation coefficients, in the ranges [0.08,0.15]
for $S_{\rm corrected~to~1~AU}$ and [0.15,0.29] for $<B>_{\rm
corrected~to~1~AU}$.

\begin{figure}[t!]
\centerline{
\includegraphics[width=0.4\textwidth, clip=]{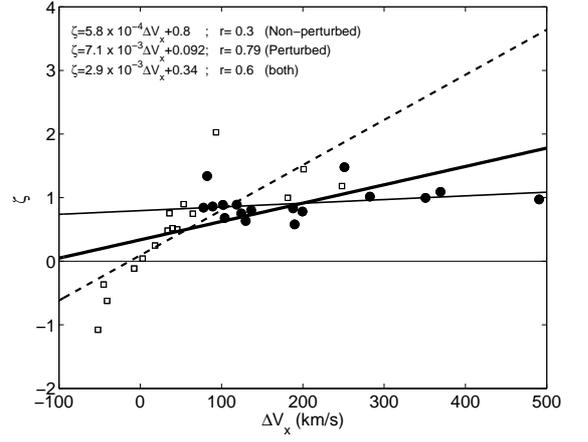}}
\caption{Perturbed and non-perturbed MCs have a remarkably different
behavior of $\zeta $ when they are plotted as a function of $\Delta
V_x$. The drawing convention is the same as in
Figs.~\ref{Fig_correlation},\ref{Fig_zeta} (A color version is
available in the electronic version).}
 \label{Fig_zeta_dV}
\end{figure}

\begin{figure}[t!]

\centerline{
\includegraphics[width=0.4\textwidth, clip=]{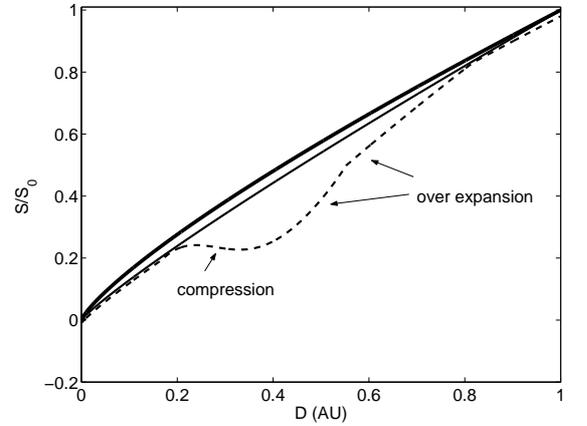}}
\caption{Cartoon of a possible evolution of the size of the MC with
the helio-distance, showing the expected global expansion (thick
solid line), an example of a non-perturbed MC (thin solid) and a
perturbed MC (dashed line).
 (A color version is available in the electronic version).}
 \label{Cartoon}
\end{figure}

\subsection{Expansion of non-perturbed MCs}
\label{E-non-overtaken}

Perturbed and non-perturbed MCs have a remarkable different behavior
of $\zeta$ as a function of $\Delta V_x$ (Fig.~\ref{Fig_zeta_dV}).
While for non-perturbed MCs the correlations have been removed, the
perturbed ones have $\zeta$ well correlated with $\Delta V_x$
($r=0.79$).

To explain the different behavior of  $\zeta$ let us take into account that
 the dependence of the size on the heliocentric distance is of the form
 (as observed from several statistical studies, Eqs.~\ref{S(D)}-\ref{Sothers(D)}):
   \begin{equation}   \label{size(D)}
   S = S_0 (D/D_0)^m \,,
   \end{equation}
where $S_0$ is the reference size at the distance $D_0$.  Its
physical origin is the approximate pressure balance between the MC
and the surrounding SW \citep[][ see
Sect.~\ref{E-origin}]{Demoulin09}. This physical driving of the
expansion is expected to induce a smooth expansion so that the size
of an individual MC closely follows Eq.~(\ref{size(D)}). Then, for
non-perturbed MCs, we can differentiate Eq.~(\ref{size(D)}) with
time to derive the expansion velocity $\Delta V_x$:
   \begin{equation}   \label{dS_dt}
   \Delta V_x \approx \frac{dS}{dt}
              \approx \frac{dS}{dD} \frac{dD}{dt}
              \approx m \frac{S}{D} ~V_c  \,.
   \end{equation}
Then, the non-dimensional expanding rate of Eq.~(\ref{zeta}) is:
   \begin{equation}   \label{zeta_not_overtaken}
   \zeta_{\rm  non-perturbed} = \frac{\Delta V_{x}}{\Delta t} \frac{D}{V_c^2}
                             \approx m  \,.
   \end{equation}

This implies that $\zeta$ is independent of the size and the
velocity of the MC as well as its distance from the Sun and its
global expansion rate $\Delta V_{x}$, in agreement with our results
(Figs.~\ref{Fig_zeta},\ref{Fig_zeta_dV}).  It also implies that the
velocity measurements across a given MC permits to estimate the
exponent $m$ in Eq.~(\ref{size(D)}). Indeed the values of $m$
deduced from the statistical study of MC size versus distance
(Eq.~\ref{S(D)}) are in agreement with the mean value found
independently with the velocity measurements
(Fig.~\ref{Fig_zeta_dV}, Table~\ref{table_res}).

\subsection{Expansion of perturbed MCs} 
\label{E-overtaken}

   For perturbed MCs, the estimation of $\zeta$ both from the size evolution
with distance and from the measured velocity still gives consistent
estimations (Eq.~\ref{S(D)}, Table~\ref{table_res}). However, the
mean value of $\zeta$ is lower ($\zeta \approx m \approx 0.45$) than
for non-perturbed MCs ($\zeta \approx m \approx 0.8$).  More
important, $\zeta$ has a much larger spread for perturbed than
non-perturbed MCs (by a factor $\approx 4$ larger). Such a larger
dispersion of $\zeta$ is expected to be the result of the variable
physical parameters (such as ram pressure) of the overtaking
streams, and also because the interaction is expected to be in a
different temporal stage for different perturbed MCs (ranging from
the beginning to the end of the interaction period).

 The effect of an overtaking stream is simply to compress the MC (at first thought), so MCs perturbed
 by this effect are expected to have a lower $\zeta$ than non-perturbed MCs. This is true in average,
 but there is a significant fraction (5/16) of perturbed MCs that are in fact expanding faster than
  the mean of non-perturbed MCs. The largest $\zeta$ is also obtained for a perturbed MC.
  Moreover, $\zeta$ for perturbed MCs still has a good correlation with $\Delta V_{x}$, opposite to
 the result obtained for non-perturbed MCs  (Fig.~\ref{Fig_zeta_dV}). Why do perturbed MCs have these properties?

When an MC is overtaken by a fast SW stream, it is compressed by the
ram, plasma and magnetic pressure of the overtaking stream, so its
size increases less rapidly with solar distance (than without
interaction). If the interaction is strong enough, this can indeed
stop the natural expansion and create an MC in compression, as in
the 3 cases present in Table~\ref{table_res}, where $\Delta
V_{x}<0$. A sketch of such evolution is given in Fig.~\ref{Cartoon}.
However, this interaction will not last for a long period of time
since the overtaking stream can overtake the flux rope from both
sides.  As the total pressure in the back of the MC decreases, the
expansion rate of the MC increases.  Indeed, its expansion rate
could be faster than the typical one for non-perturbed MCs, as
follows. The compression has provided an internal pressure that is
stronger than the surrounding SW total pressure. Then, when the
extra pressure of the overtaking stream has significantly decreased,
the MC has an over-pressure compared to the surrounding SW, so it
expands faster than usual i. e., an overexpansion, see
\citet{Gosling95}. Indeed, the flux rope is expected to evolve
towards the expected size that it would have achieved without the
overtaken SW flow.

So, depending on which time the MC is  observed in the interaction
process, a perturbed MC can expand slower or faster than without
interaction (Fig.~\ref{Cartoon}). This explains the dispersion of
$\zeta$ found for perturbed MCs, but also the presence of some MCs
with faster expansion than usual.

In the case of perturbed MCs, their sizes still follow
Eq.~(\ref{size(D)}) on average (as shown by Eq.~\ref{S(D)}), but it
has no meaning to apply this law locally to a given MC. In
particular, we cannot differentiate Eq.~(\ref{size(D)}) with time to
get an estimation of the local expansion velocity of a perturbed MC
(so we cannot write Eq.~\ref{dS_dt}). Rather we can use
Eq.~(\ref{size(D)}) only to have an approximate size $S$ in the
expression of $\zeta$:
   \begin{equation}   \label{zeta_overtaken}
   \zeta_{\rm perturbed} = \frac{\Delta V_{x}}{S}\frac{D}{V_c}
               \approx \frac{\Delta V_{x}{D_0}^m D^{1-m}}{S_0 V_c} \,.
   \end{equation}
The dependence of $D^{1-m} \approx D^{0.5}$ is relatively weak, but
still present (Fig.~\ref{Fig_zeta}a). $\zeta$ also has a dependence
on $V_c$, but the range of $V_c$ within the studied perturbed group
is very limited to derive a reliable dependence from the
observations (Fig.~\ref{Fig_zeta}b), and, moreover, there is a
dependence on $S_0$ that we cannot quantify. It remains that the
strongest correlation of $\zeta$ is with $\Delta V_{x}$
(Fig.~\ref{Fig_zeta_dV}).   Indeed, for perturbed MCs, $\zeta$
strongly reflects their local expansion behavior, so that $\zeta$ is
strongly correlated to $\Delta V_{x}$.

\begin{figure}[t!]
\includegraphics[width=0.5\textwidth, clip=]{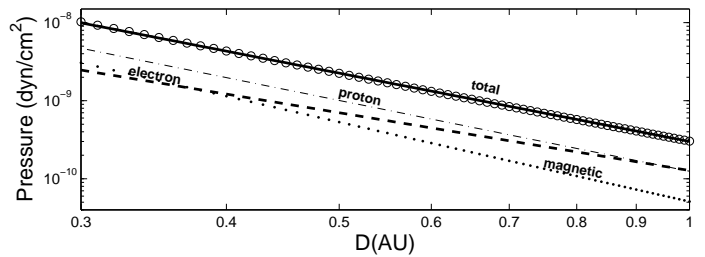}
\caption{Log-log plots of the SW total pressure and its components
as a function of the solar distance.  The solid line shows the least
square straight line fitted to the total pressure as computed from
the points marked with circles. The total pressure in the SW
decreases as $P_{sw}(D) = P_0 D^{-2.9}$ (A color version is
available in the electronic version).}
 \label{Fig_Psw}
\end{figure}

\subsection{Physical origin of MC expansion} 
\label{E-origin}

The main driver of MC expansion is the rapid decrease of the total
SW pressure with solar distance \citep{Demoulin09}. Other effects,
such as the internal over-pressure, the presence of a shock, as well
as the radial distribution and the amount of twist within the flux
rope have a much weaker influence on the expansion. This result was
obtained by solving the MHD equations for flux ropes having various
magnetic field profiles, and with ideal MHD or fully relaxed states
(minimizing magnetic energy while preserving magnetic helicity).
Within the typical SW conditions, they have shown that any
force-free flux rope will have an almost self similar expansion, so
a velocity profile almost linear with time as observed by a
spacecraft crossing an MC (e.g.
Figs.~\ref{Fig_not_overtaken},\ref{Fig_overtaken}). They also relate
the normalized expansion rate $\zeta$ to the exponent $n_P$ of the
total SW pressure as a function of the distance $D$ to the Sun
(defined by $p \propto D^{-n_P}$) as $\zeta \approx n_P/4$.

Here we further test the above theory with the MCs analyzed in this
paper, by comparing the value obtained for $\zeta$ with the value of
$n_P$ obtained from previous studies of the SW.

According to \cite{Mariani90}, from fitting a power law to observations
of the field strength in the inner heliosphere according to $B = B_0 (D/D_0)^{-n_B}$,
a global decay law is obtained with $n_B=1.6 \pm 0.1$ ($B_0=3.8 \pm 0.2$ nT at $D_0 = 1$AU) from Helios 1, and $n_B=1.8 \pm 0.1$ ($B_0=3.3 \pm 0.2$ nT at $D_0 = 1$AU) from Helios 2.

For the proton density ($N_p = N_{p,0} (D/D_0)^{-n_N}$) we consider
a density of $N_0=7\pm 4$~cm$^{-3}$ at 1~AU (averaging slow and fast
SW according to \citealp{Schwenn06}) and $n_N=2$ (corresponding to
the 2D expansion for the stationary SW with constant radial
velocity).

According to \cite{Schwenn06} and \cite{Totten95}, it is possible to
represent a typical dependence of the proton temperature ($T_p$)
upon $D$ as approximately $T_p = T_{p,0} (D/D_0)^{-n_{T_p}}$ with
$n_{T_p} =1.0 \pm 0.1$ ($T_{p,0}=(1.3\pm 1.0) \times 10^5$K at $D_0
= 1$AU).

For electron temperature ($T_e = T_{e,0} (D/D_0)^{-n_{T_e}}$) we
follow \cite{Marsch89}, in particular their results for the ranges
of velocities (300-500)~km/s to better represent the typical
conditions of the SW. For the velocity range (300-400)~km/s,
\cite{Marsch89} found $T_{e,0}=(1.3 \pm 0.4) \times 10^5$ K and
$n_{T_e}=0.5 \pm 0.1$; for the velocity range (400-500)~km/s,
$T_{e,0}=(1.4 \pm 0.4) \times 10^5$ K and $n_{T_e}=0.4 \pm 0.1$. For
electrons we then consider mean temperature averaged over these two
ranges of SW speeds.

The partial pressures (magnetic, proton, and electron) are shown in
Fig.~\ref{Fig_Psw}.  Neglecting the small effect of the $\alpha$
particles, the total pressure in the SW ($P_{SW}$) is:
$P_{SW} = P_B + P_p + P_e$. We then propose that the total pressure
also follows a power law ($P_{SW} = P_0 D^{-n_P}$) and fit this
power law to $P_{SW}(D_i)$. Then, we fit $P_{SW}$ with a power law
($P_0 D^{-n_P}$). The sum of different power laws is generally not a
power law, however in the present case we still find a total
pressure which is very close to a power law since the magnetic and
plasma pressures have similar exponents. This also implies that the
exponent found, $n_P = 2.91 \pm 0.31$, has a low sensitivity to the
plasma $\beta$ and the relative pressure contribution of the
electrons and protons.

Using the result of \citet{Demoulin09} that $\zeta \approx n_P/4$
for force-free flux rope, we found $\zeta = 0.73 \pm 0.08$, in full
agreement, within the error bars, with $\zeta$ found from velocity
measurement in non-perturbed MCs (Table~\ref{table_res}). This
further demonstrates that MC expansion is mainly driven by the
decrease of the surrounding SW total pressure with solar distance.
The main departure from this global evolution is due to the presence
of overtaking flows.

\section{Summary and discussion of main results}
\label{summary}
MCs have a specific magnetic configuration, forming flux ropes which
expand in all directions, unlike the almost 2D expansion of the
surrounding SW. But how fast do they expand? Is the expansion rate
specific for each MC or is there a common expansion rate? What is
the role of the surrounding SW? Finally, what is the main driver of
such 3D expansion?

In order to answer these questions we have re-analyzed a significant
set of MCs observed by both Helios spacecrafts. In order to better
define the MC extension, we first analyzed the magnetic data,
finding the direction of the flux rope axis, and then we rotated the
magnetic data in the MC local frame.  This step is important to
separate the axial and ortho-axial field components since they have
very different spatial distributions in a flux rope, and since we
then can use the magnetic flux conservation of the azimuthal
component as a constraint on the boundary positions (e.g., as done
in \cite{Dasso06,Steed08}). Then, in the local MC frame, we can
better define the boundaries of the MC.

The observed velocity profile typically has a linear variation with
time, with a larger velocity in front than in the back of the MC.
This is a clear signature of expansion. On top of this linear trend,
fluctuations of the velocity are relatively weak, with the most
noticeable exception occurring when an overtaking fast stream is
observed in the back of the MC. Such fast flow can enter the MC,
removing the linear temporal trend. We consider these overtaken MCs
in a separate group as perturbed MCs (Fig.~\ref{Fig_overtaken}). We
exclude from the analysis the MCs where the overtaking flow was
extending more than half the MC size and MCs where the data gaps
were too large. The remaining MCs are classified as non-perturbed,
even if some of them have weak perturbations in their velocity
profiles. These perturbations are filtered by considering only the
major part of the velocity profile where the profile is almost
linear with time (Fig.~\ref{Fig_not_overtaken}).

The group of non-perturbed MCs has a broad and typical range of
sizes and magnetic field strengths ($\approx [0.1,0.75]$ AU and
$\approx [7,25]$ nT respectively when rescaled at 1~AU). They also
have a broad range of expansion velocities ($\approx [80,500]$
km/s). Such a range of expansion velocity cannot be explained by the
range of observing distances, ($D$ in $[0.3,1]$~AU), since the
expansion velocity decreases only weakly within this range of $D$.
However, we found that the expansion velocity is proportional to the
MC size. By further analyzing the correlation between the observed
expansion velocity and other measured quantities, such as the MC
velocity, we empirically defined a non-dimensional expansion
coefficient $\zeta$ (Eq.~\ref{zeta}). For the non-perturbed MCs,
$\zeta$ is independent of all the other characteristics of the MCs
(such as size and field strength). Moreover, this empirical
definition of $\zeta$, obtained by removing the correlation in the
data between the expansion rate and other quantities, finally
defines the same $\zeta$ quantity as the one defined from
theoretical considerations by \citet{Demoulin08}. We conclude that
$\zeta$ characterizes the expansion rate of non-perturbed MCs.

For the non-perturbed MCs, we found that $\zeta$ is confined to a
narrow interval: $0.91\pm 0.23$.  This is consistent with the result
obtained at 1~AU for a set of 26 MCs observed by Wind and ACE
\citep{Demoulin08}. Indeed, we found that $\zeta$ is independent of
solar distance (within $[0.3,1]$ AU) in the Helios MCs.

What is the origin of this common expansion rate of MCs?
\citet{Demoulin09} have shown theoretically that the main origin of
MC expansion is the decrease of the total SW pressure with solar
distance $D$. With a SW pressure decreasing as $D^{-n_P}$, they
found that $\zeta \approx n_P/4$ independently of the magnetic
structure of the flux rope forming the MC. In the present work, we
re-analyzed the total SW pressure variation with $D$, revising
previous studies that also analyzed Helios data
\citep{Mariani90,Schwenn06,Totten95,Marsch89}. We found $n_P=2.91
\pm 0.31$, which implies $n_P/4=0.73 \pm 0.08$, in agreement with
our estimation of $\zeta$ from the measured velocity in MCs. We then
confirm that the fast decrease of the total SW pressure with solar
distance is the main cause of the MC expansion rate.

For MCs overtaken by a fast SW stream (or by another flux rope on
its back, e.g. \cite{Dasso09}), we minimize its importance in the
estimation of the MC expansion rate by considering only the part of
the velocity profile which is nearly linear with time. Still, the
mean computed $\zeta$ for perturbed MCs is significantly lower than
the mean value for non-perturbed MCs, showing that the overtaking
flows have a more global effect on MCs (than the part where the
velocity profiles significantly depart from the linear temporal
behavior).  A lower expansion rate is a natural consequence of the
compression induced by the overtaking flow.

More surprising, some perturbed MCs are found to expand faster
(larger $\zeta$) than non-perturbed MCs.  We conclude that such MC
are probably observed after the main interaction phase with the
overtaking flow, so that they expand faster than usual in order to
sustain an approximate pressure balance with the surrounding SW
(Fig.~\ref{Cartoon}). More precisely, as the overtaking flow
disappears from the back of the MC, the MC is expected to tend
towards the size that it would have reached without the interaction
with the fast stream. Since it was compressed, it is expanding
faster than usual to return to its expected size in a normal SW.

\section{Conclusions} 
\label{conclusion}
Our present results confirm and extend our previous work on the
expansion of MCs. The non-dimensional expansion factor $\zeta$
(Eq.~\ref{zeta}) gives a precise measure of the expanding state of a
MC.  In particular, it removes the size effect which could give the
false impression that parcels of fluid in large MCs are expanding
faster. The value of $\zeta$ in non-significantly perturbed MCs is
clustered in a narrow range, independent of their magnetic
structure, but determined by the pressure gradient of the
surrounding SW.  The mean value of $\zeta$ is also nearly the
exponent of the solar distance for the MC size (determined from the
analysis of MCs at different solar distances, section
~\ref{M-MC-size}).

However, for perturbed MCs, $\zeta$ has a much broader range, a
result linked to its proportionality to the local expansion
velocity. So for perturbed MCs, $\zeta$ is a measure of the local
expansion rate and of the importance of the overtaking stream (i.e.,
a quantification of the influence of the MC/stream interaction on
the expansion of the MC).

 Finally for non-perturbed MCs, $\zeta$ is independent of the solar
  distance in the inner heliosphere. Presently we do not know how far
   this result extends to larger distances, even if it is an expected
   result as long as the flux ropes still exist. This will be the subject of a future study.

\begin{acknowledgements}
We thank the referee for reading carefully, and improving the
manuscript.
The authors acknowledge financial support from ECOS-Sud
through their cooperative science program (N$^o$ A08U01).
This work was partially supported by the Argentinean grants:
UBACyT X425 and PICTs 2005-33370 and 2007-00856 (ANPCyT).
S.D. is member of the Carrera del Investigador Cien\-t\'\i fi\-co, CONICET.
A.M.G. is a fellow of Universidad de Buenos Aires. M.E.R is a fellow of CONICET.
\end{acknowledgements}

\bibliographystyle{aa}
\bibliography{mc}
\IfFileExists{\jobname.bbl}{}
{\typeout{}
\typeout{****************************************************}
\typeout{****************************************************}
\typeout{** Please run "bibtex \jobname" to optain}
\typeout{** the bibliography and then re-run LaTeX}
\typeout{** twice to fix the references!}
\typeout{****************************************************}
\typeout{****************************************************}
\typeout{}
}


\newpage

{\large \bf Next pages: Color version of some figures (for the electronic version) }

\newpage

\setcounter{figure}{0}
\begin{figure*}[t!]
\centerline{
\includegraphics[width=0.4\textwidth, clip=]{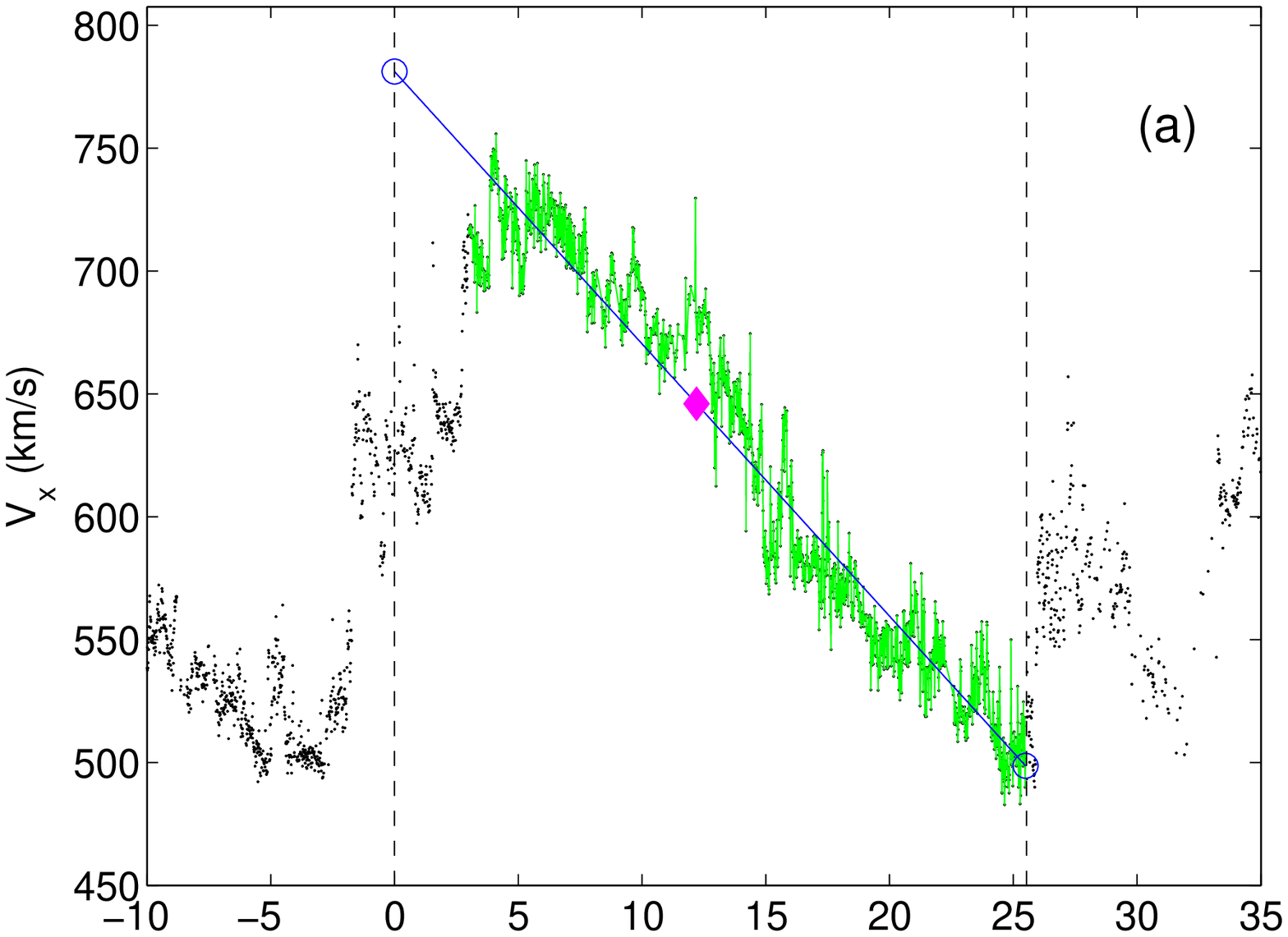}
\includegraphics[width=0.4\textwidth, clip=]{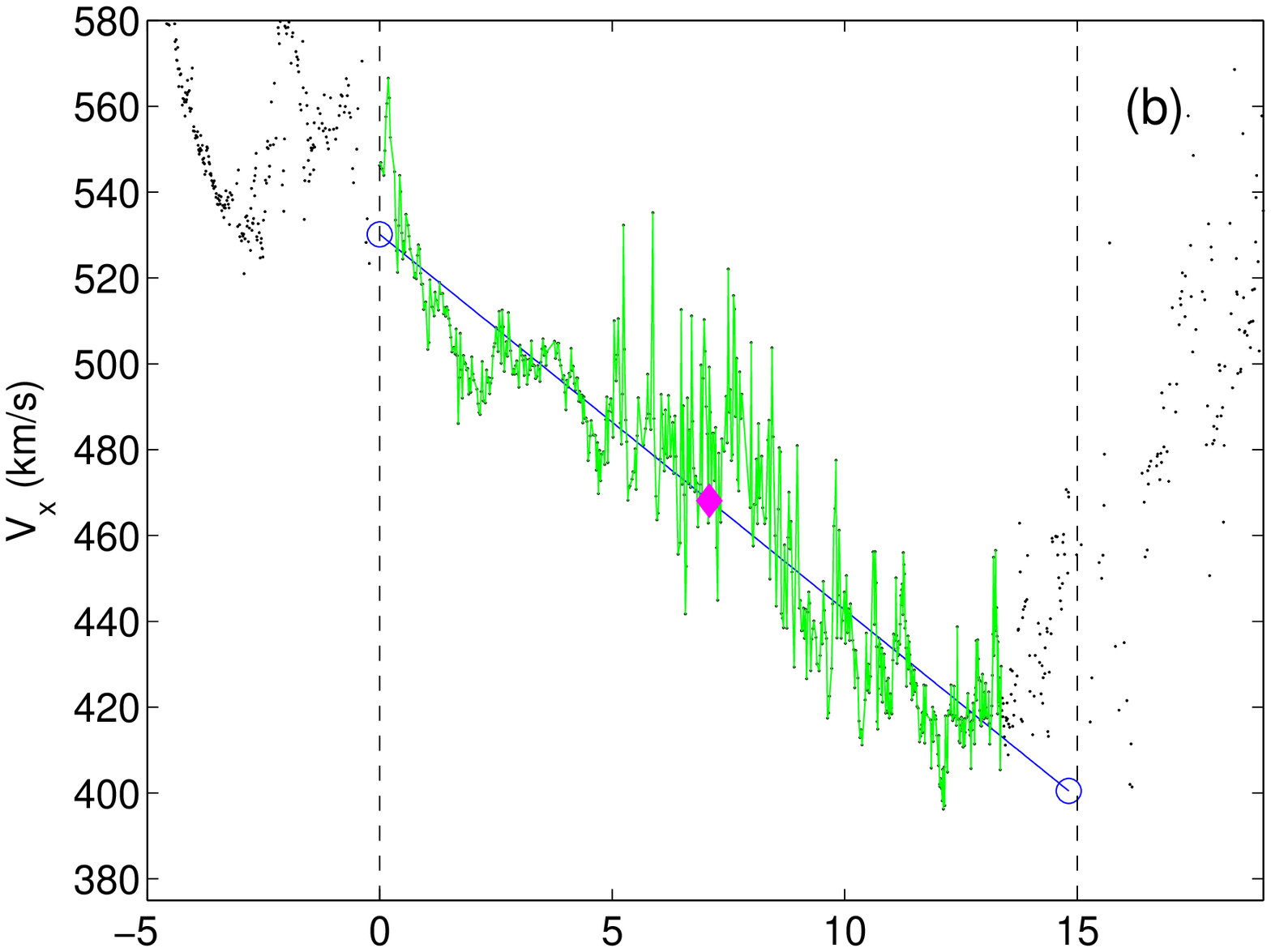}}
\centerline{\includegraphics[width=0.4\textwidth,clip=]{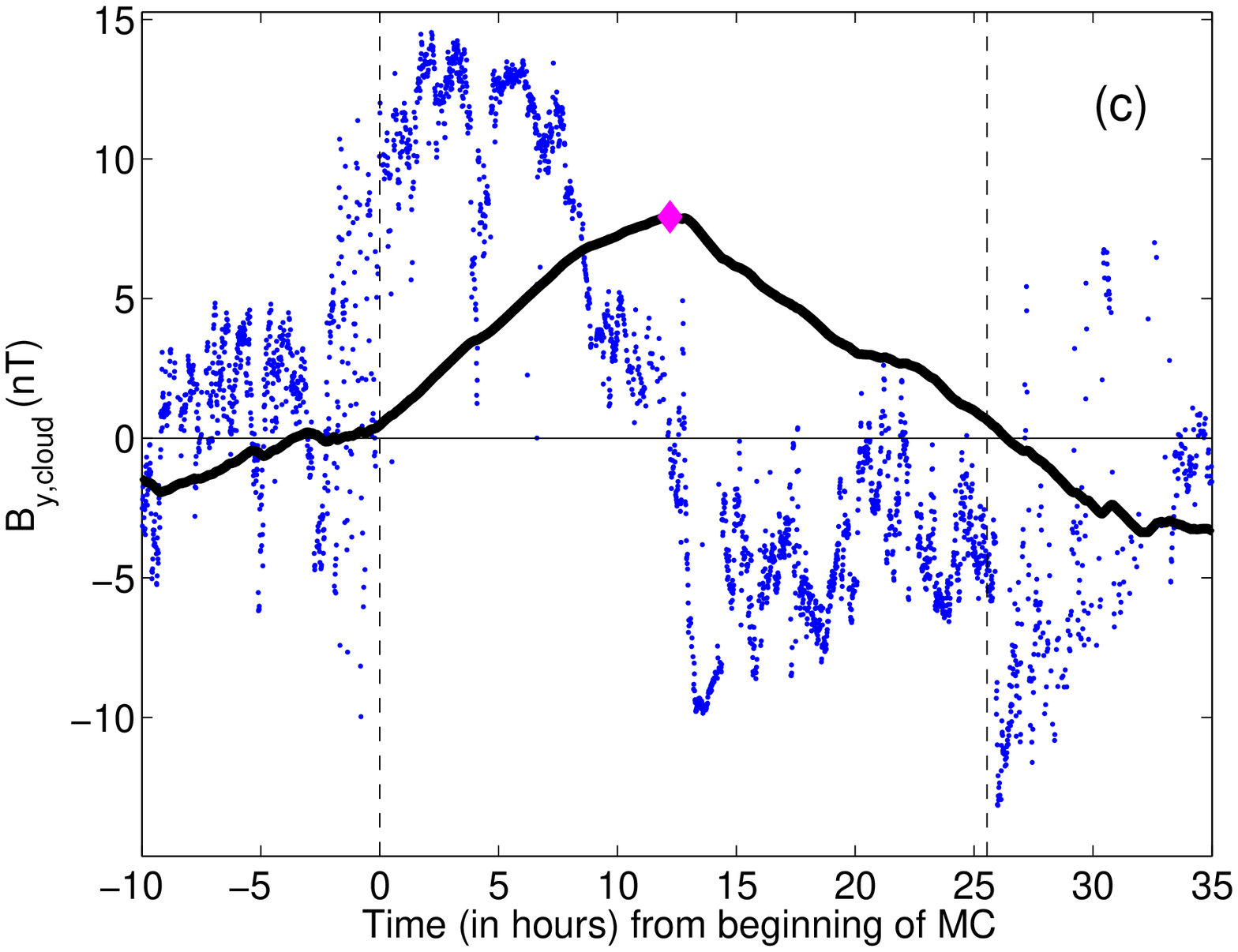}
\includegraphics[width=0.4\textwidth, clip=]{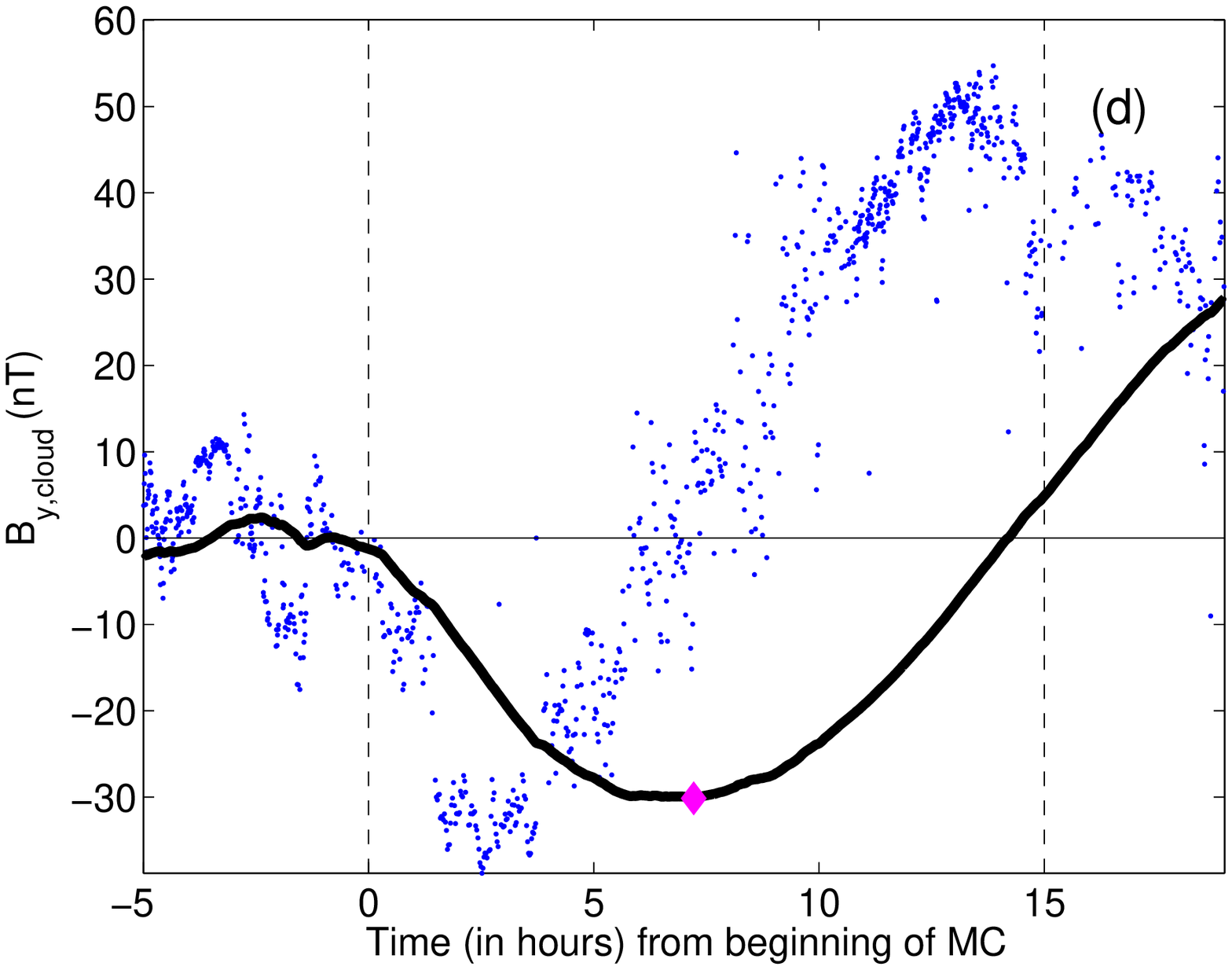}}
\caption{Examples of two analyzed MCs that are not significantly
perturbed by a fast flow. The MC center was observed at 07-Jan-1975
10:39 and 04-Mar-1975 21:37 UT, for panels (a,c) and (b,d),
respectively.   The vertical dashed lines define the MC boundaries.
     (a,b) $V_{x}$ is the observed velocity component in the radial direction from the Sun,
expressed in km per second. The straight line is the linear least
square fit of the velocity in the time interval where an almost linear trend is present
(where the observations are presented as a solid line). The linear fitting is extrapolated to the borders of the MC, which are marked with circles.
  (c,d) $B_{y}$ is the magnetic field component, in nT, both orthogonal to the MC axis and to the spacecraft trajectory,  while the solid line represents $F_{y}$, which is the accumulated flux of $B_{y}$ (Eq.~\ref{Byaccumul}). The extremum of $F_y$ (proxy of the cloud center) is indicated with diamonds.}
 \label{Fig_not_overtaken}

\end{figure*}
\begin{figure*}[t!]
\centerline{
\includegraphics[width=0.4\textwidth, clip=]{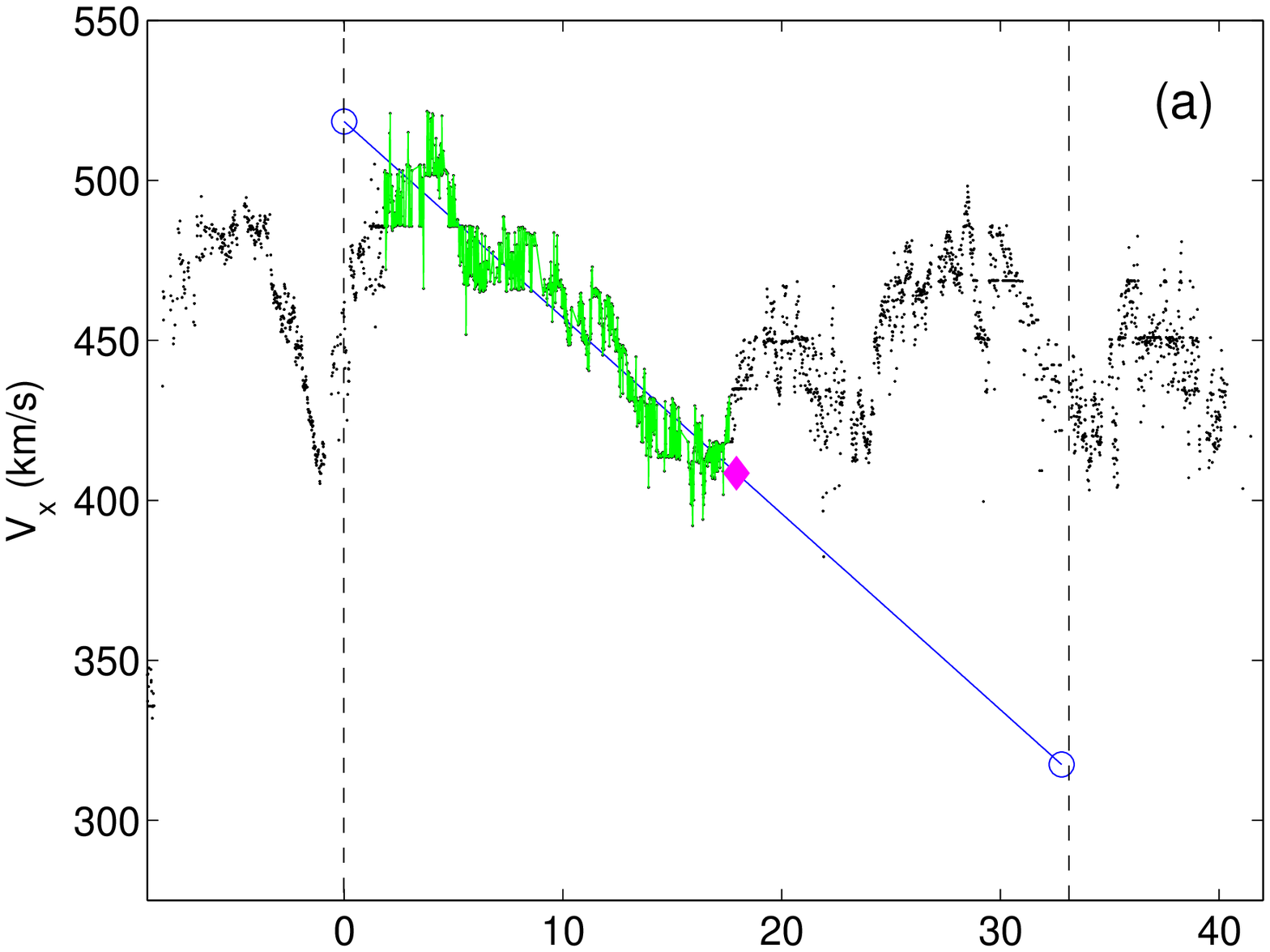}
\includegraphics[width=0.4\textwidth, clip=]{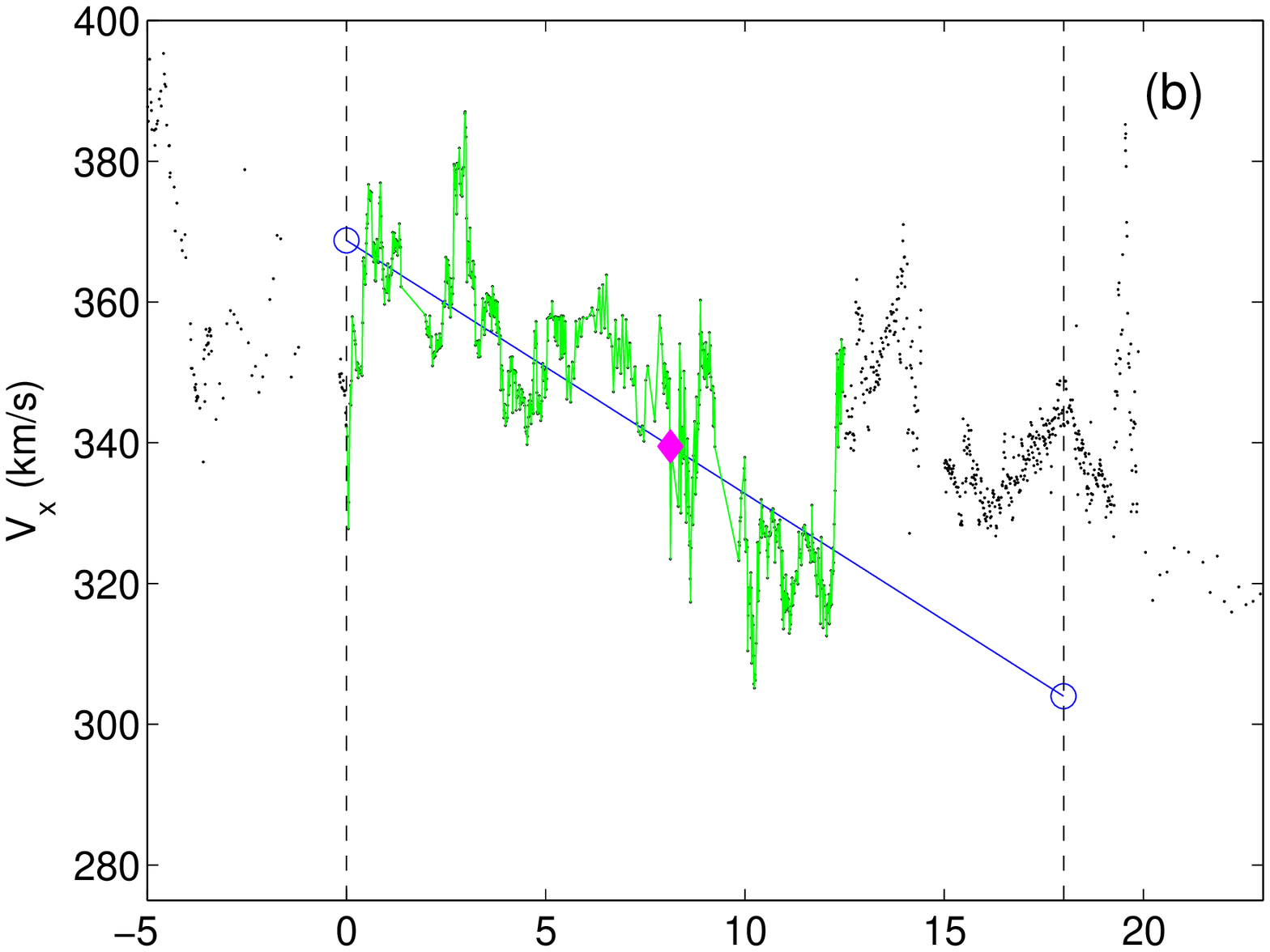}}
\centerline{\includegraphics[width=0.4\textwidth,clip=]{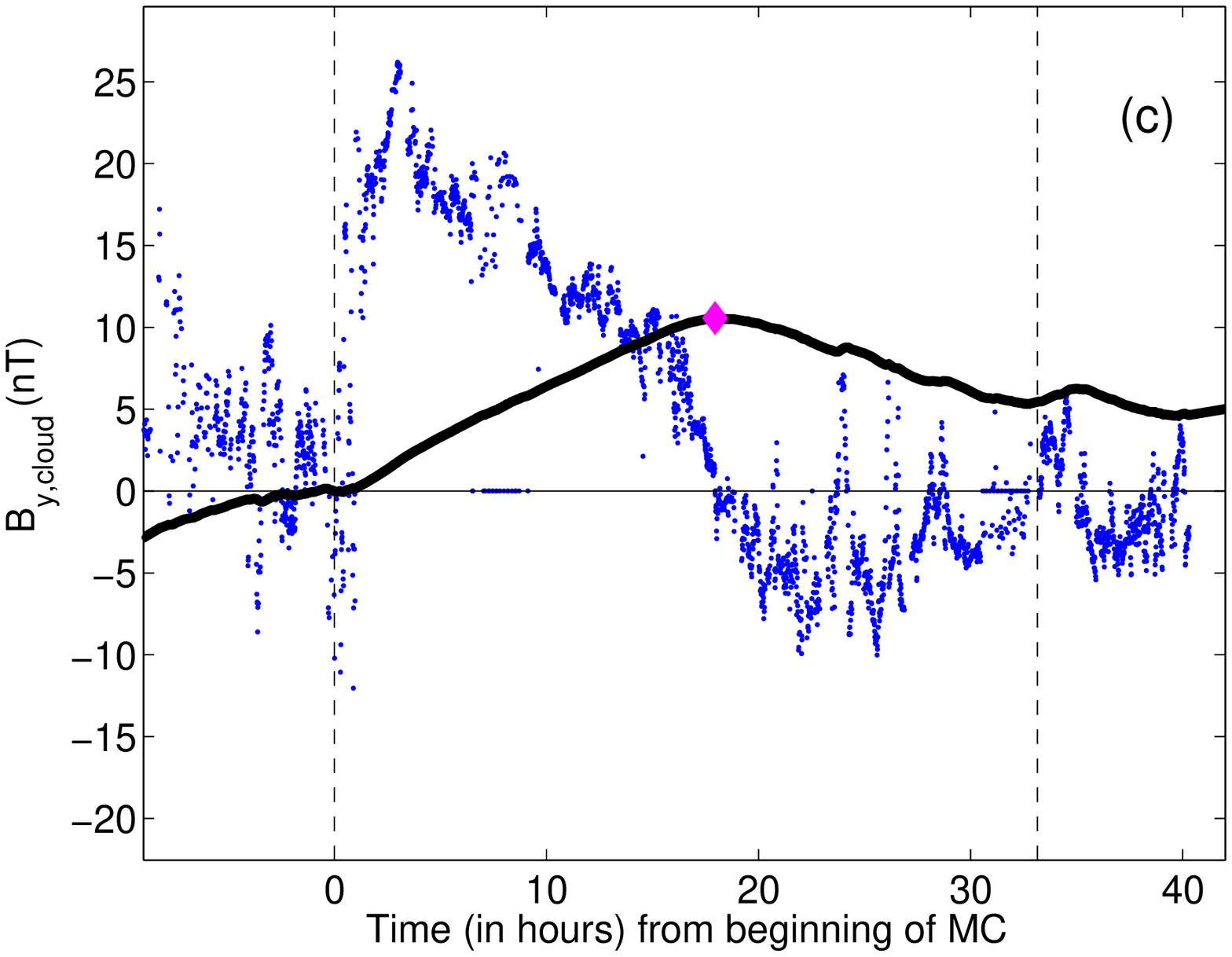}
\includegraphics[width=0.4\textwidth, clip=]{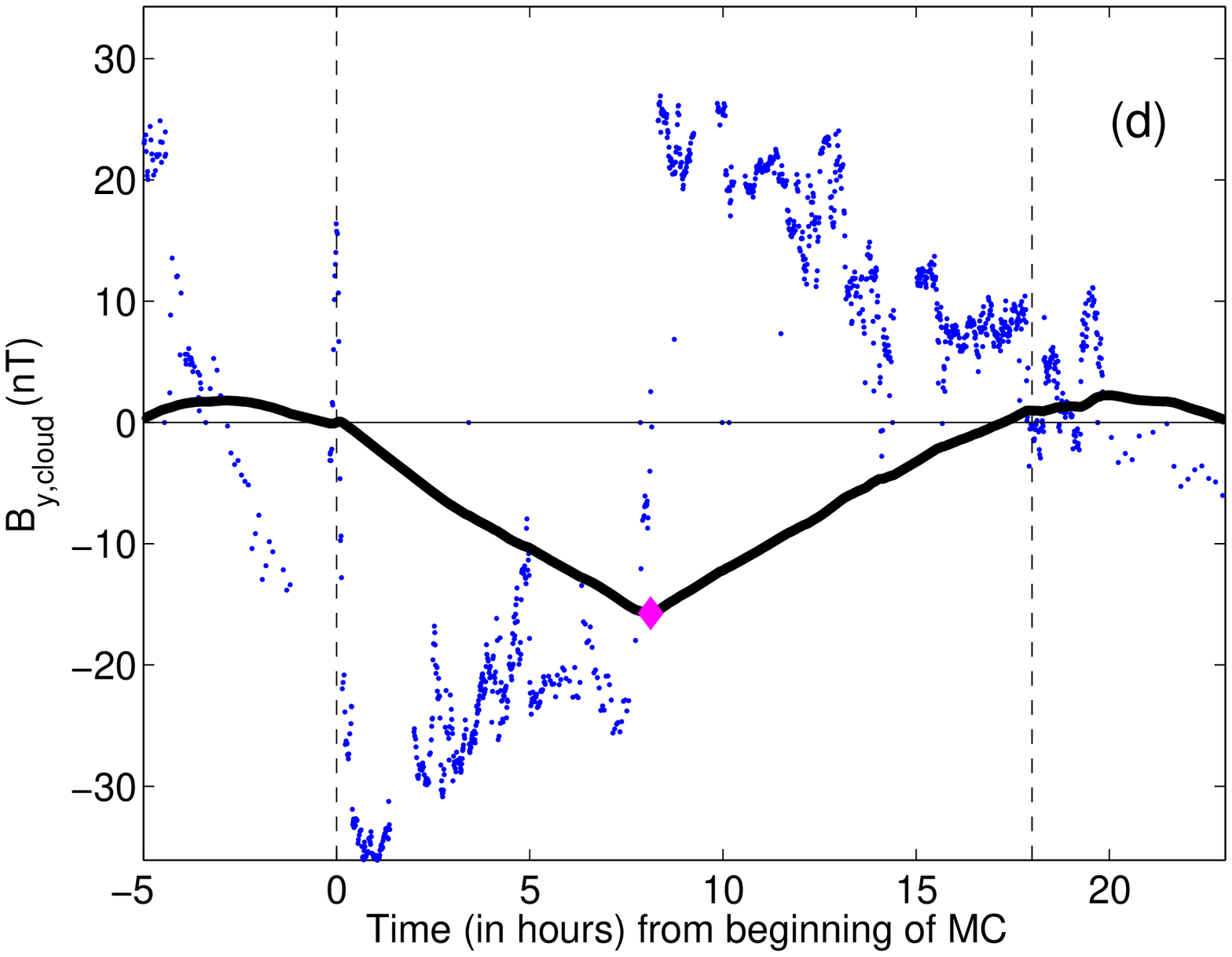}}
\caption{Examples of two analyzed MCs that are perturbed by a fast
flow as seen in the upper panels. The MC center was observed at
30-Jan-1977 03:18 and 23-Jun-1980 12:25 UT, for panels (a,c) and
(b,d), respectively. The external perturbations enter in a
significant part of the MCs. The same quantities, as in
Fig.~\ref{Fig_not_overtaken}, are shown.}
\end{figure*}
\begin{figure*}[t!]
\centerline{
\includegraphics[width=0.4\textwidth, clip=]{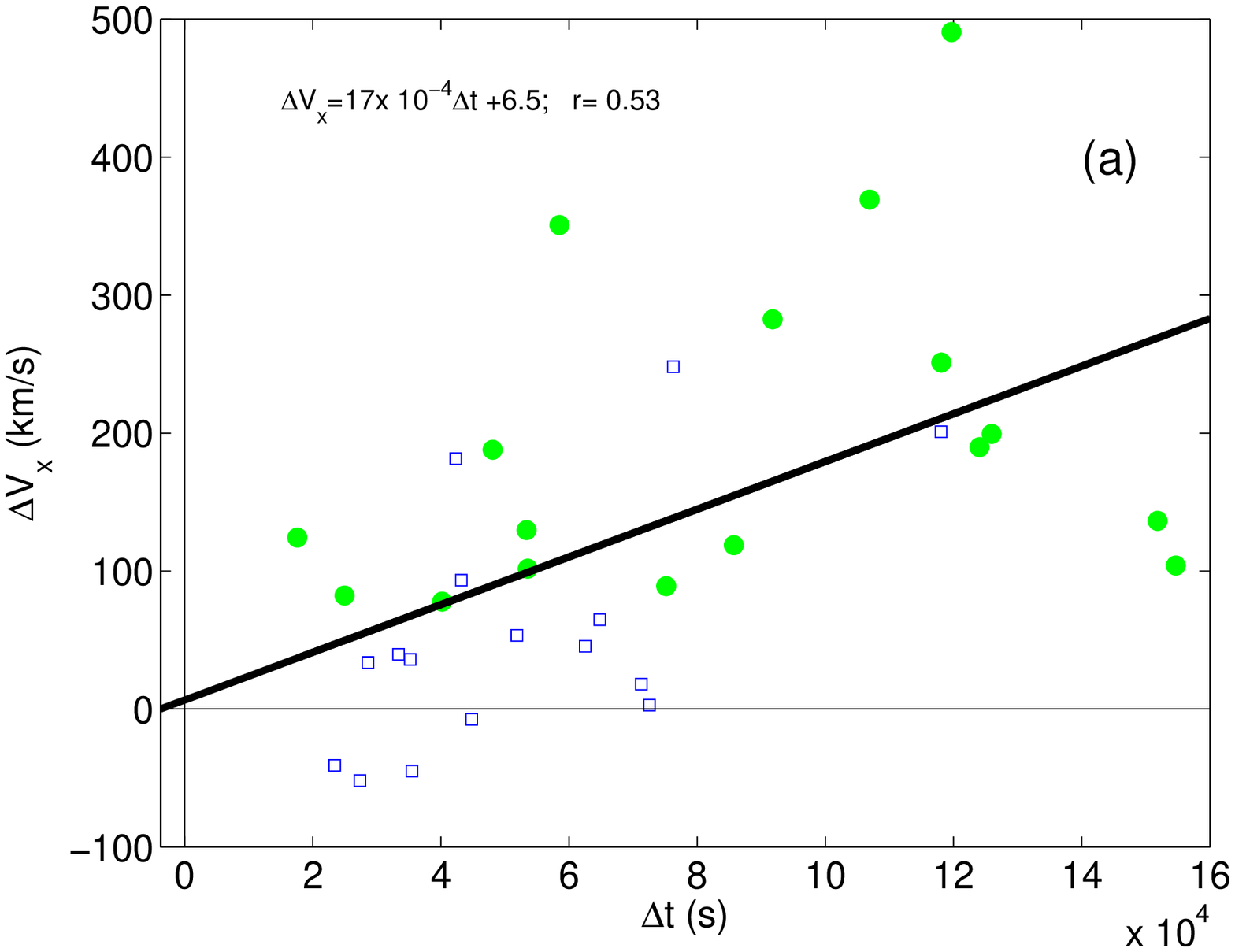}
\includegraphics[width=0.4\textwidth, clip=]{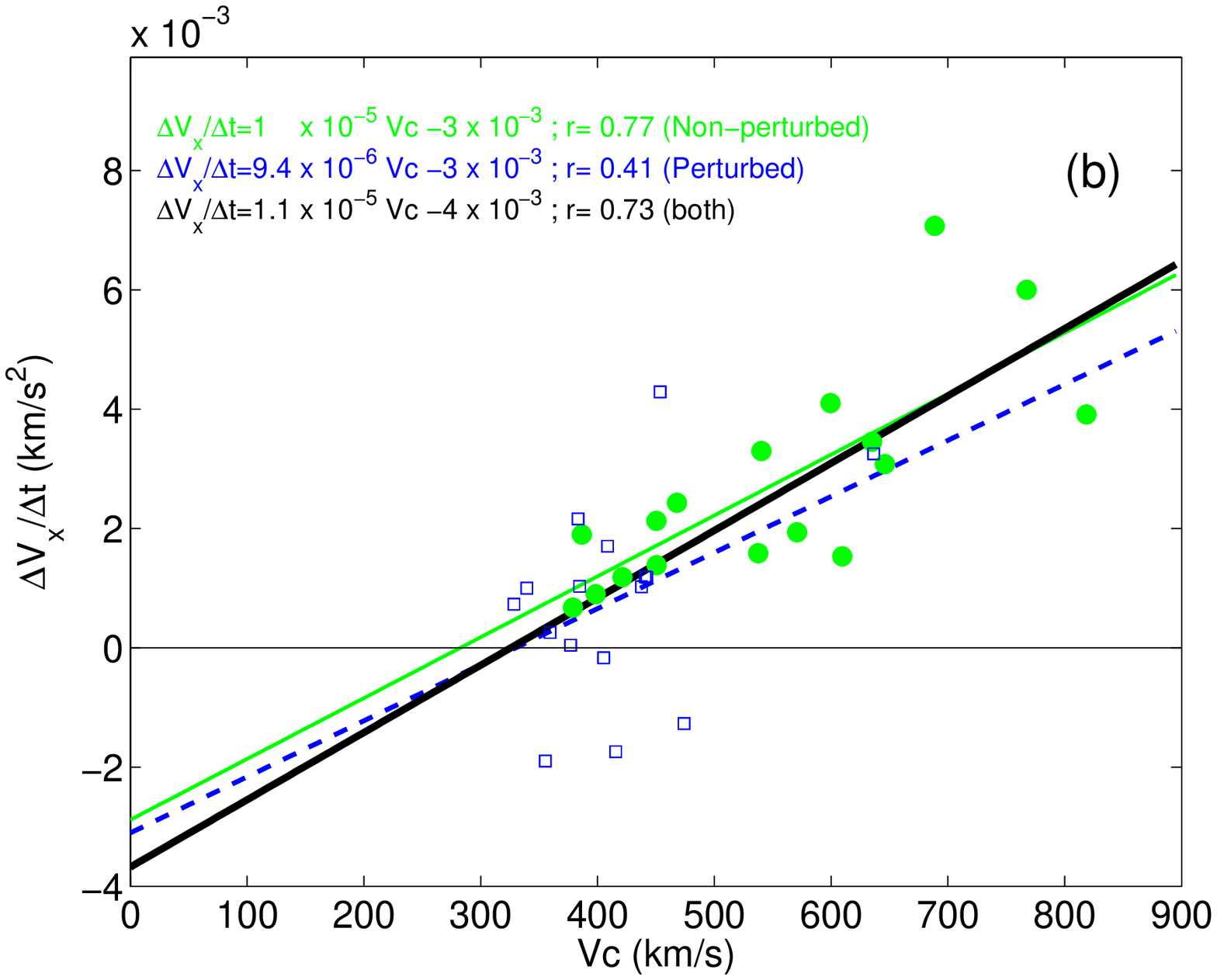}}
\centerline{\includegraphics[width=0.4\textwidth,clip=]{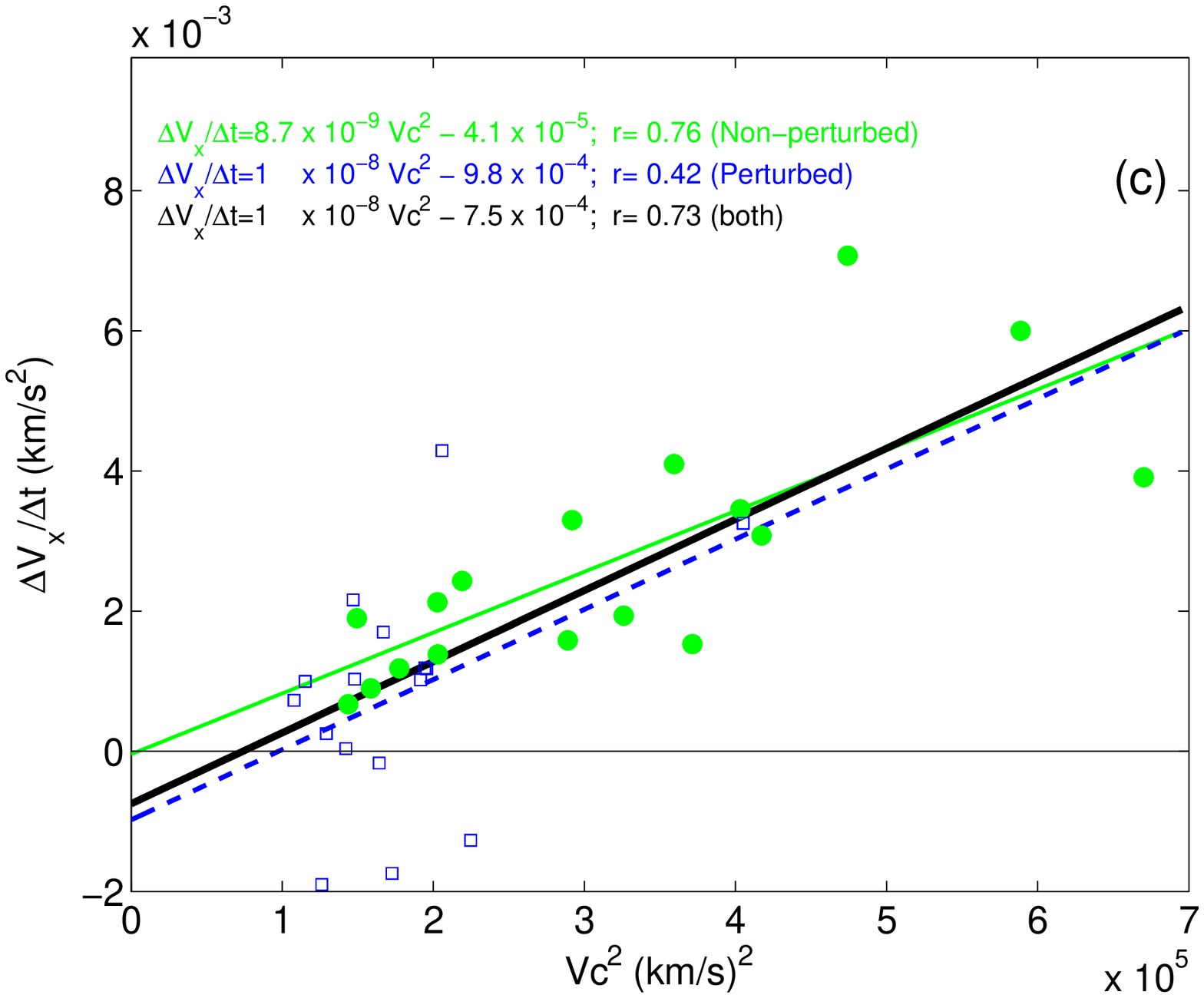}
\includegraphics[width=0.4\textwidth, clip=]{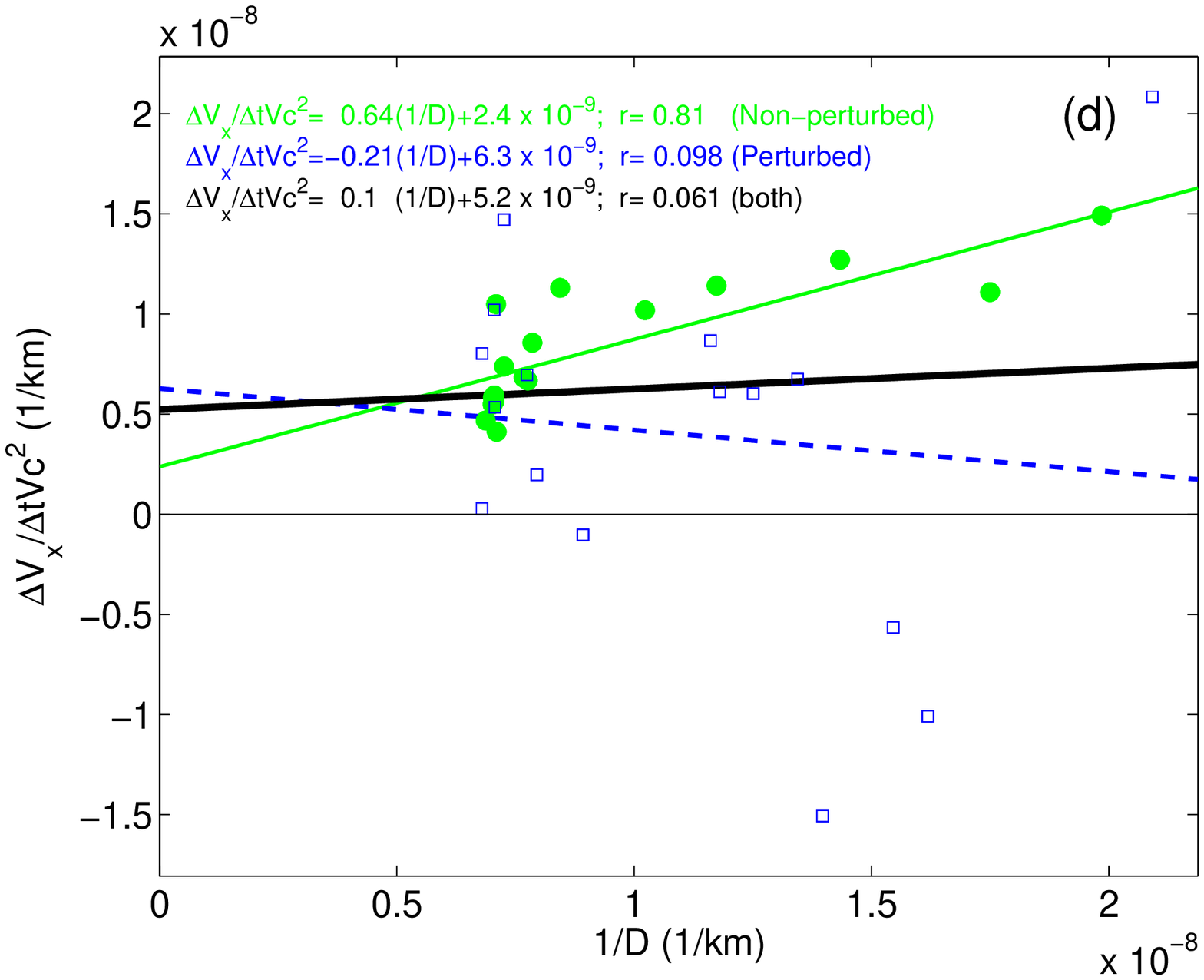}}
\caption{The panels a-d show the correlation analysis between
proxies for MC expansion with different physical quantities. MCs are
separated in two groups: perturbed (empty square symbol) and
non-perturbed (filled circle symbol). The straight lines are the
result of a least square fit for perturbed (dashed line),
non-perturbed (thin continuous line), and for both set of MCs in the
list of events shown in Table~\ref{table_res} (thick continuous
line). $\Delta V_x $ is defined by Eq.~(\ref{dV}), $\Delta t =t_{\rm
out}- t_{\rm in}$, $V_c$ is the velocity of the MC center (or at the
closest approach distance), and $D$ is the distance to the Sun.  The
fitted values and the obtained correlation coefficient ($r$) are
included as insets, considering the different groups. More
significant correlation is present for the non-perturbed cases. }
\end{figure*}
\begin{figure*}[t!]
\centerline{
\includegraphics[width=0.4\textwidth, clip=]{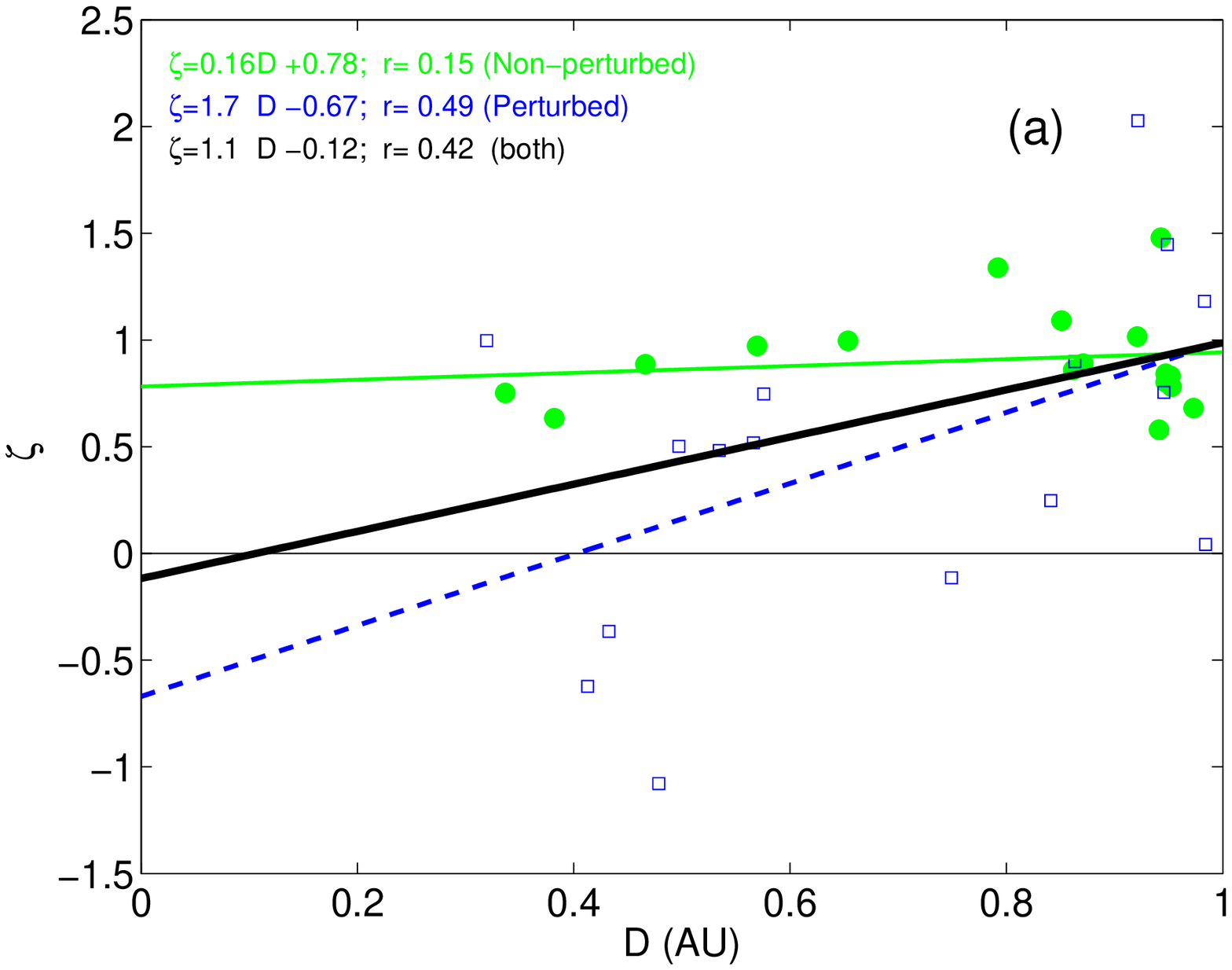}
\includegraphics[width=0.4\textwidth, clip=]{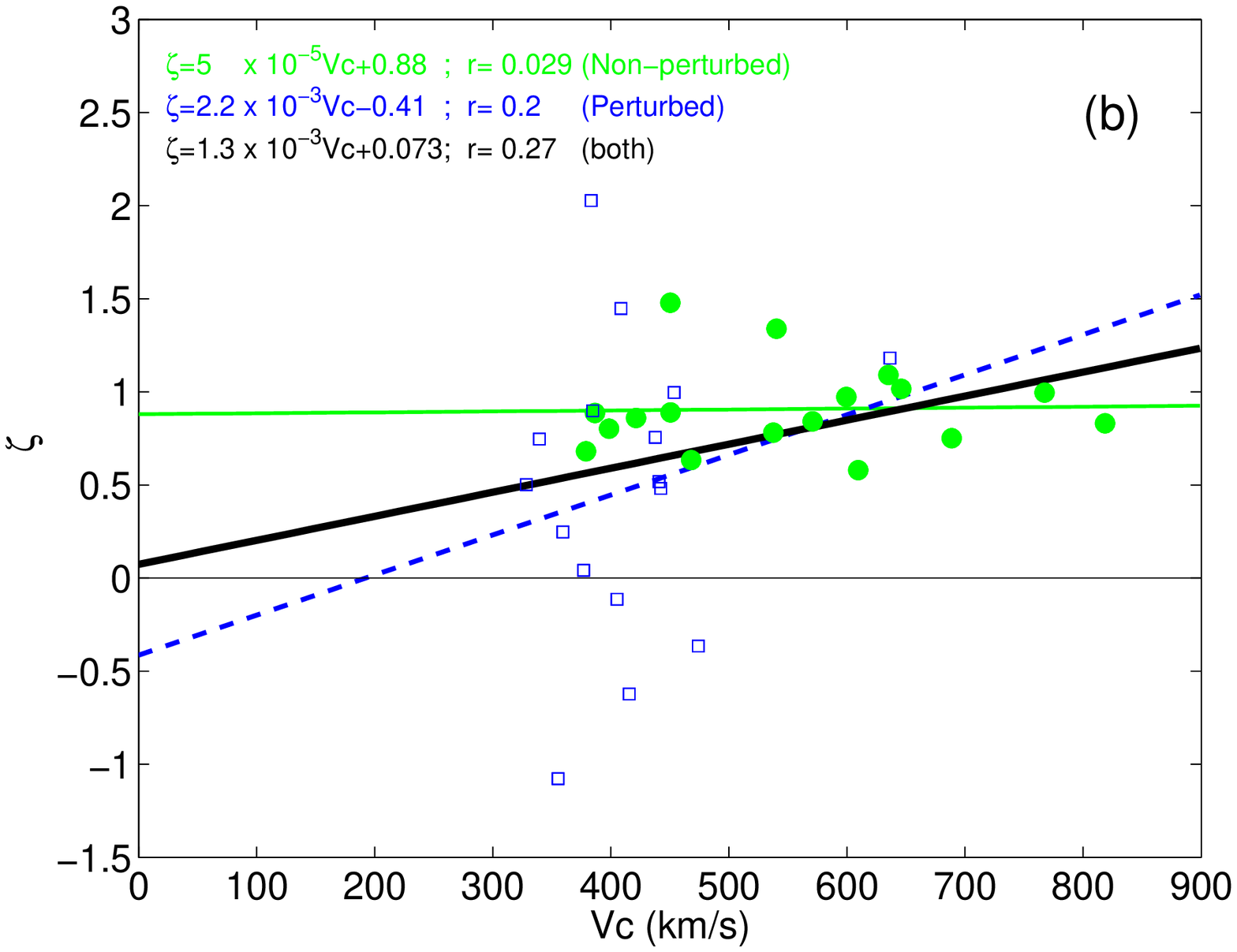}}
\centerline{\includegraphics[width=0.4\textwidth,clip=]{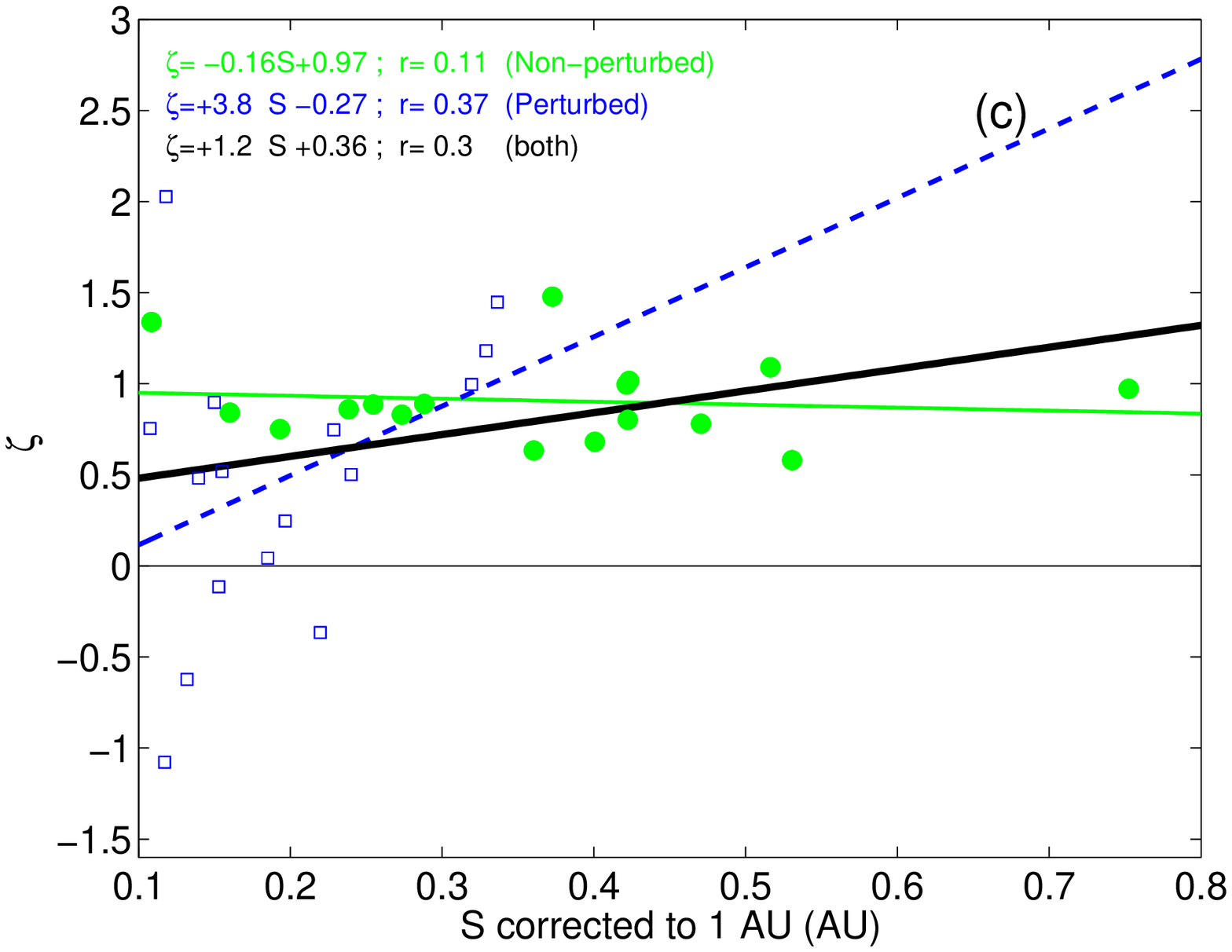}
\includegraphics[width=0.4\textwidth, clip=]{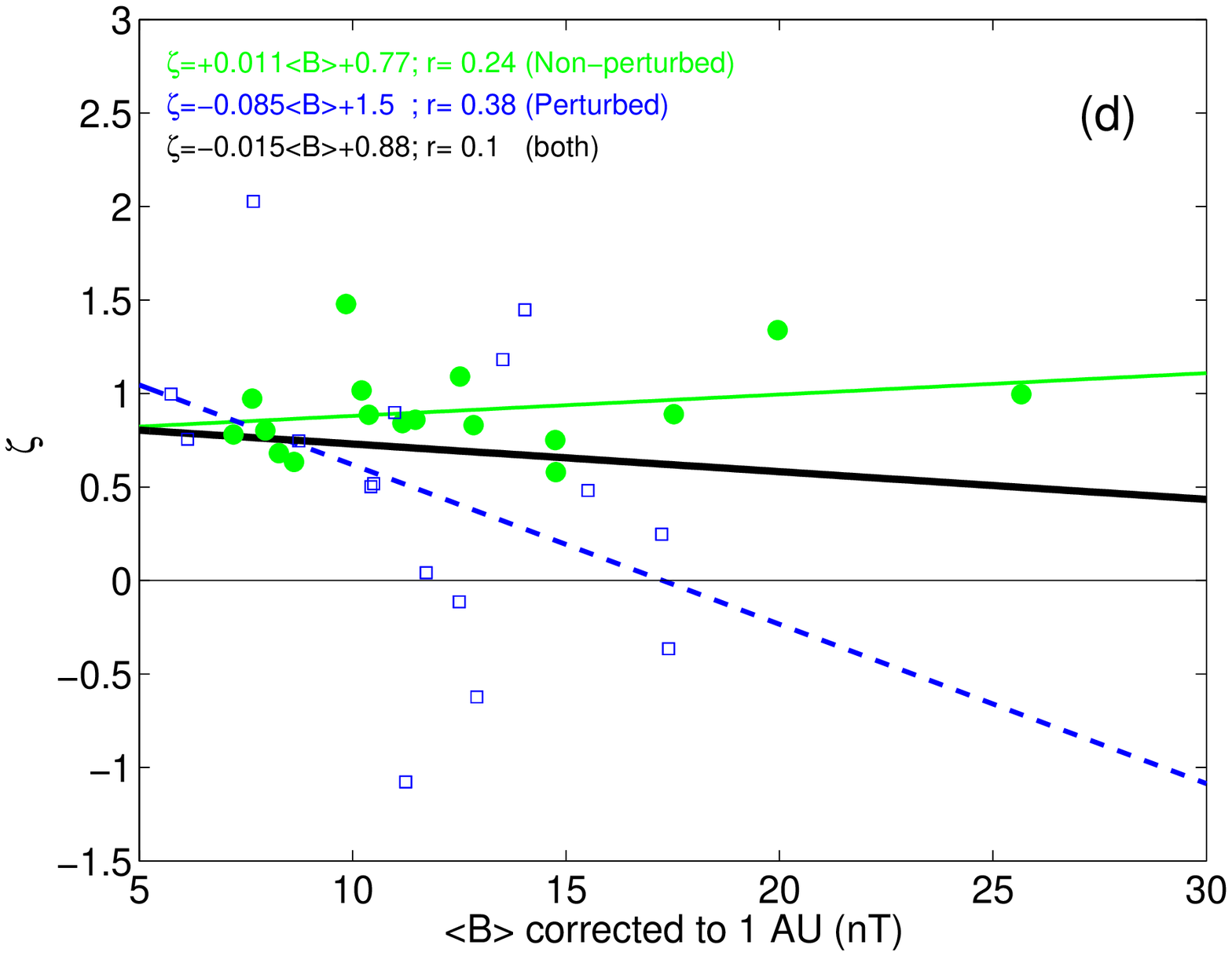}}
  \caption{Panels a-d show the correlation analysis that tests for the dependence
   of the non-dimensional expansion factor $\zeta$ (Eq.~\ref{zeta}) as a function of other parameters.
Perturbed and non-perturbed MCs are represented as in
Fig.~\ref{Fig_correlation}. $S_{\rm corrected~to~1~AU}$ and
$<B>_{\rm corrected~to~1~AU}$ are normalized to 1 AU using the size
and field strength dependence on the distance, according to the
relationship given in Eq.~(\ref{S+Bcorrected}). }
\end{figure*}
\begin{figure}[t!]
\centerline{
\includegraphics[width=0.4\textwidth, clip=]{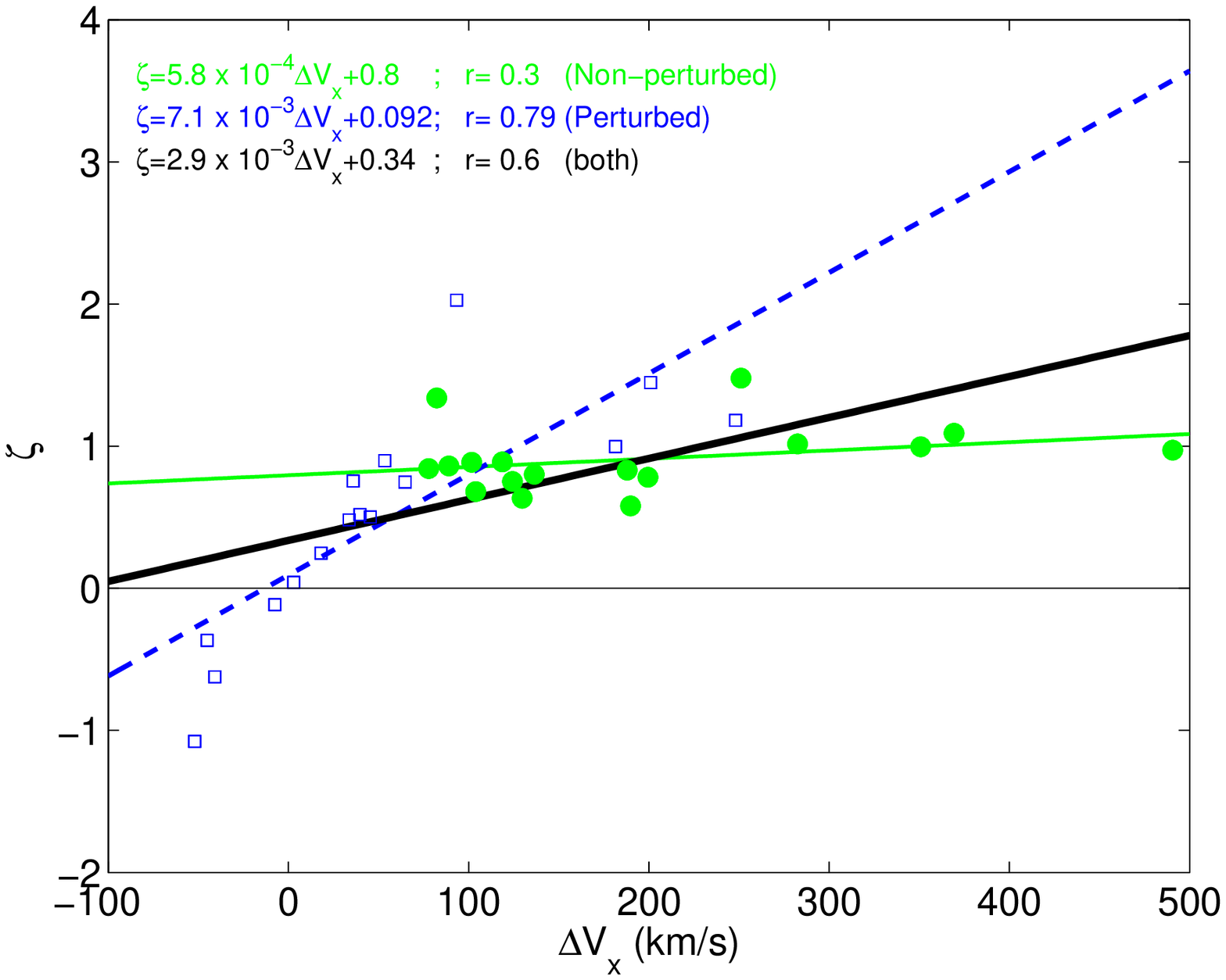}}
\caption{Perturbed and non-perturbed MCs have a remarkably different
behavior of $\zeta $ when they are plotted as a function of $\Delta
V_x$. The drawing convention is the same as in
Figs.~\ref{Fig_correlation},\ref{Fig_zeta}.}
\end{figure}
\begin{figure}[t!]

\centerline{
\includegraphics[width=0.4\textwidth, clip=]{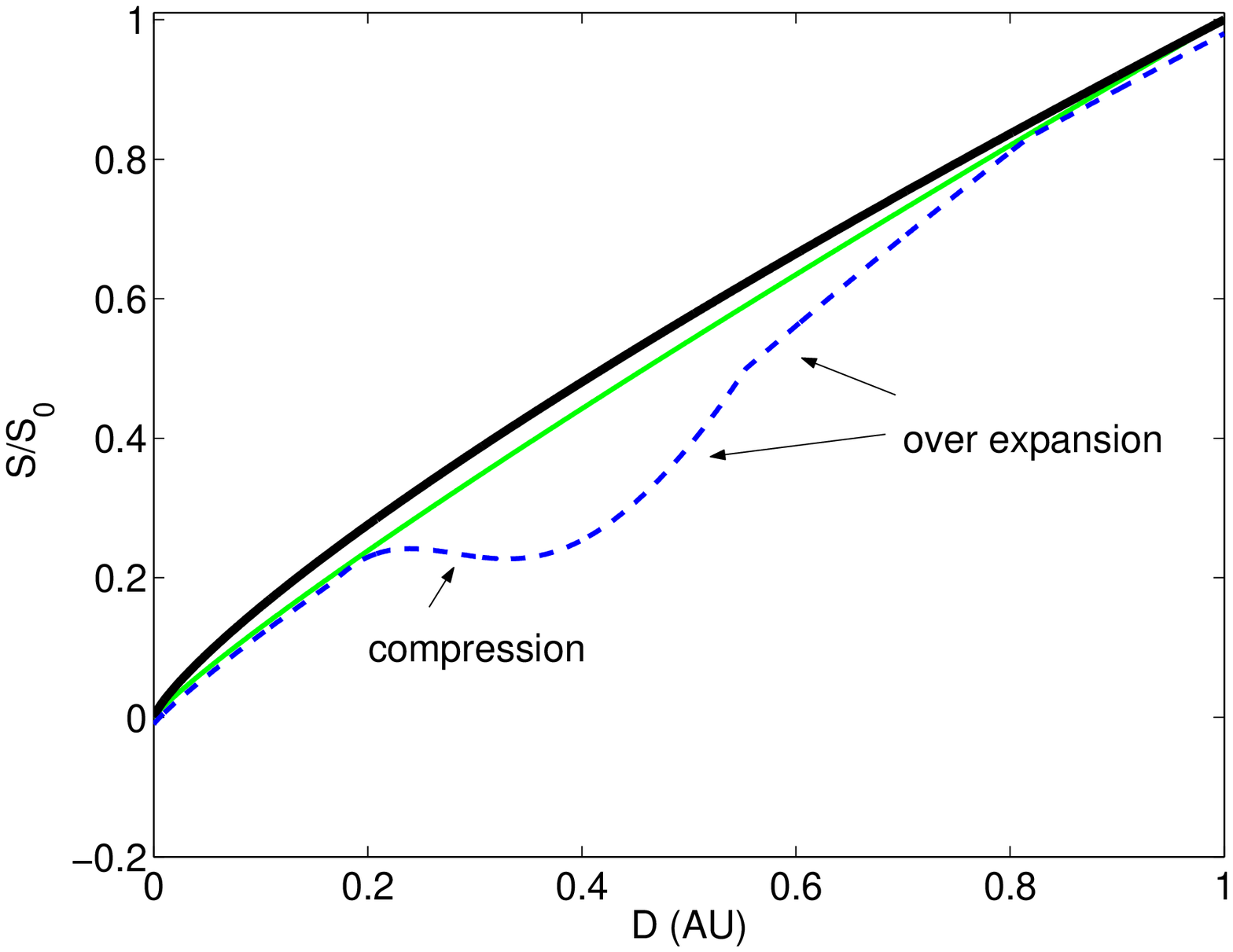}}
\caption{Cartoon of a possible evolution of the size of the MC with
the helio-distance, showing the expected global expansion (thick
solid line), an example of a non-perturbed MC (thin solid) and a
perturbed MC (dashed line). }
\end{figure}

\begin{figure}[t!]
\includegraphics[width=0.5\textwidth, clip=]{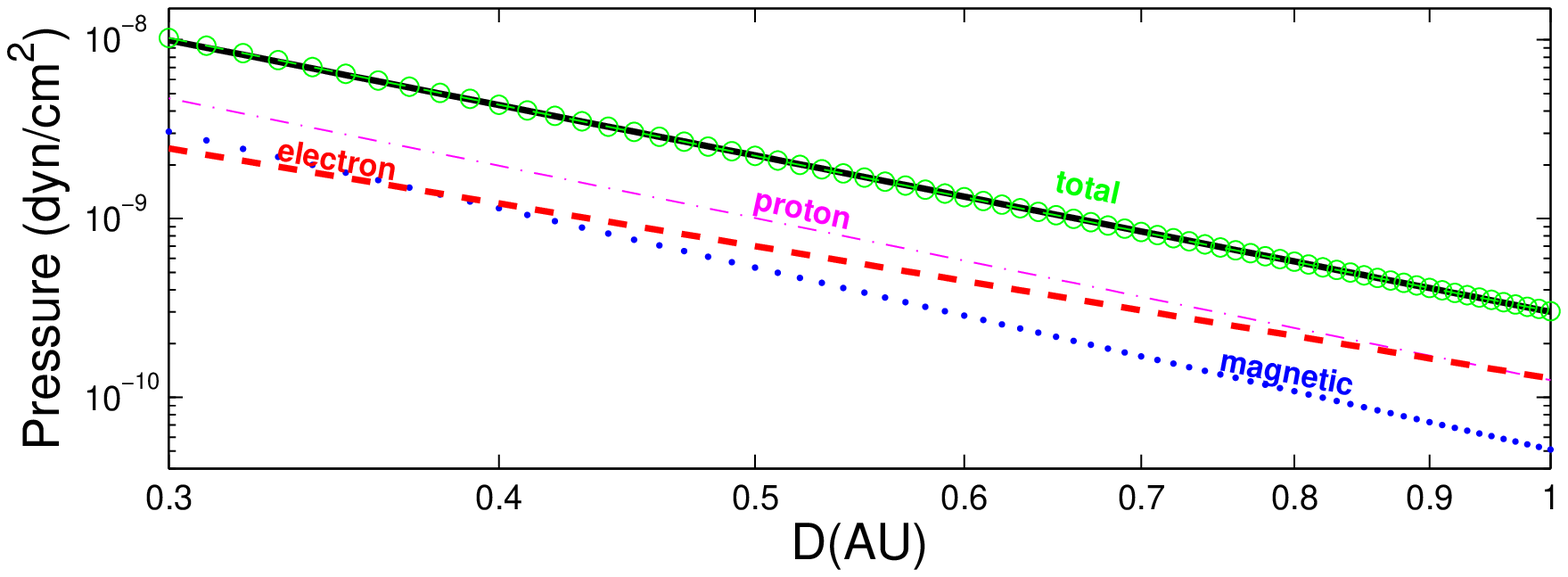}
\caption{Log-log plots of the SW total pressure and its components
as a function of the solar distance.  The solid line shows the least
square straight line fitted to the total pressure as computed from
the points marked with circles. The total pressure in the SW
decreases as $P_{sw}(D) = P_0 D^{-2.9}$.}
\end{figure}

\end{document}